\RequirePackage[2020-02-02]{latexrelease}
\documentclass[prb,twocolumn,showpacs,amysymb]{revtex4}
\usepackage{graphicx}
\usepackage{tabularx}
\usepackage{dcolumn}
\usepackage{color}
\usepackage{bm}

\begin{document}

\title{
Spin-orbit coupling effects in altermagnets: Interplay of weak spin and orbital
ferromagnetism with relativistic splitting of electron states}

\author{L. M. Sandratskii$^{1,2}$\footnote{lsandr3591@gmail.com}, K. Carva$^1$, V. M. Silkin$^{2,3,4}$}
\affiliation{$^1$Faculty of Mathematics and Physics, Charles University, 12116 Prague,
Czech Republic\\
$^2$Donostia International Physics Center (DIPC), Paseo de Manuel Lardizabal 4, E-20018 San Sebasti\'an, Spain\\
$^3$
Departamento de Pol\'{\i}meros y Materiales Avanzados: F\'{\i}sica,
Qu\'{\i}mica y Tecnolog\'{\i}a, Facultad de Ciencias
Qu\'{\i}micas, Universidad del Pa\'{\i}s Vasco (UPV-EHU), Apdo. 1072,
E-20080 San Sebasti\'an, Spain\\
$^4$IKERBASQUE, Basque Foundation for Science, 48011 Bilbao, Spain\\
}

\begin{abstract}
The novel class of collinear compensated magnets, dubbed altermagnets, has attracted
immense research attention by the property of nonrelativistic spin splitting. More
recently, the properties of altermagnets caused by relativistic spin-orbit coupling (SOC)
became the topic of many investigations. The aim of the paper is to contribute to
a deeper understanding of the formation of relativistic effects in altermagnets.
The focus of the paper is on the phenomena of weak ferromagnetism (WFM) and
relativistic modification of the electron states
including the interplay of these phenomena.
The main tools of the study are various types of relativistic calculations based on
density functional theory combined with the symmetry analysis on the basis of spin space
groups. The consideration is performed on two different levels. On the first level, the atomistic
magnetic structure of weak ferromagnetic state is calculated and discussed.
Both spin and orbital atomic moments are taken into account.
We study the dependence of the WFM moment on the strength of the SOC and obtain a peculiar nonmonotonous
type of dependence. We separate the contributions of the SOC of the metal and ligand atoms to WFM.
A very interesting result is obtained in
quasisymmetry (QS) calculation where only the component of the
SOC collinear to the Néel vector is taken into account. In QS calculation
the spin WFM is absent while the orbital
WFM is present. This result reveals a principal difference in the formation of the spin
and orbital magnetic moments. On the second level, the study is focused on the properties of individual
electron states.
We introduce the notion of the magnetic structure of the electron state (MSES).
It is shown how the nonrelativistic collinear spin-MSES of both metal and ligand moments and nonrelativistic compensated
orbital-MSES of the ligand moments transform into
complex noncollinear 3D MSESs of both spin and orbital nature giving large contributions to WFM.
An important role in the formation of MSESs is played by the relativistic
splitting of the accidental spin degeneracies at general {\bf k} points filling the volume of the Brillouin zone.
The formation of the regions of avoided crossings in the relativistic band structure
is related to the nonmonotonous dependence
of the WFM moment as the function of the SOC strength. The
importance of the metal-ligand hybridization in the formation of the AM properties
is discussed and illustrated by means of the comparison of the LDA and LDA+$U$ calculations.
Most of the calculations are performed for MnTe, a prototypical material for the class of altermagnets.
\end{abstract}

\maketitle

\section{Introduction}
The new class of compensated collinear magnets dubbed altermagnets (AMs) was defined through
nonrelativistic spin splitting of electron states in the reciprocal
space \cite{Smejkal2002,Smejkal2002A}, the property promising
numerous novel applications
\cite{Smejkal2002A,Mazin2022,Bai2024}.
In a short time the class of AMs became a hot topic of intensive theoretical and experimental
studies \cite{Guo2023,Sodequist2024_2D_AM,Osumi2024,McClarty2024,Mazin2024_gossamerWFM,Kluczyk2024_WFM_Exp,Milivojevic2024,
Krempasky_MnTe_2024, Hajlaoui2024,Orlova2025,Sorn2025,Antonenko2025,Roig2025_quasisym,
Jo2025,Carmine2025,Solovyev2026,Ye2026, Smejkal2023, Liu2024, Biniskos2025,
Beida2025, Turek2022,Sandratskii2025, Lounis_review_2025,Mostovoy2025,Trama2025,Jungwirth_Newton_review2025,Jungwirth2026,Liu2026}.
Recently the features of AMs caused by the relativistic spin-orbit coupling (SOC)
attracted much research attention.
Among the relativistic properties are weak ferromagnetism (WFM), that is a weak canting of the atomic
moments of a collinear antiferromagnet \cite{Mazin2024_gossamerWFM,Kluczyk2024_WFM_Exp,Carmine2025,Solovyev2026,Milivojevic2024}, and
relativistic splitting of electron states \cite{Krempasky_MnTe_2024,Hajlaoui2024,Antonenko2025} that can have special
features in the case of AMs.

The purpose of this paper is to reach a deeper understanding of
the influence of the SOC
on the electronic structure and properties of AMs by means of the combination of various
types of the density-functional-theory (DFT) based relativistic calculations with the symmetry
analysis on the basis of spin-space groups (SSG).

\begin{figure}[t]
\includegraphics*[width=8cm]{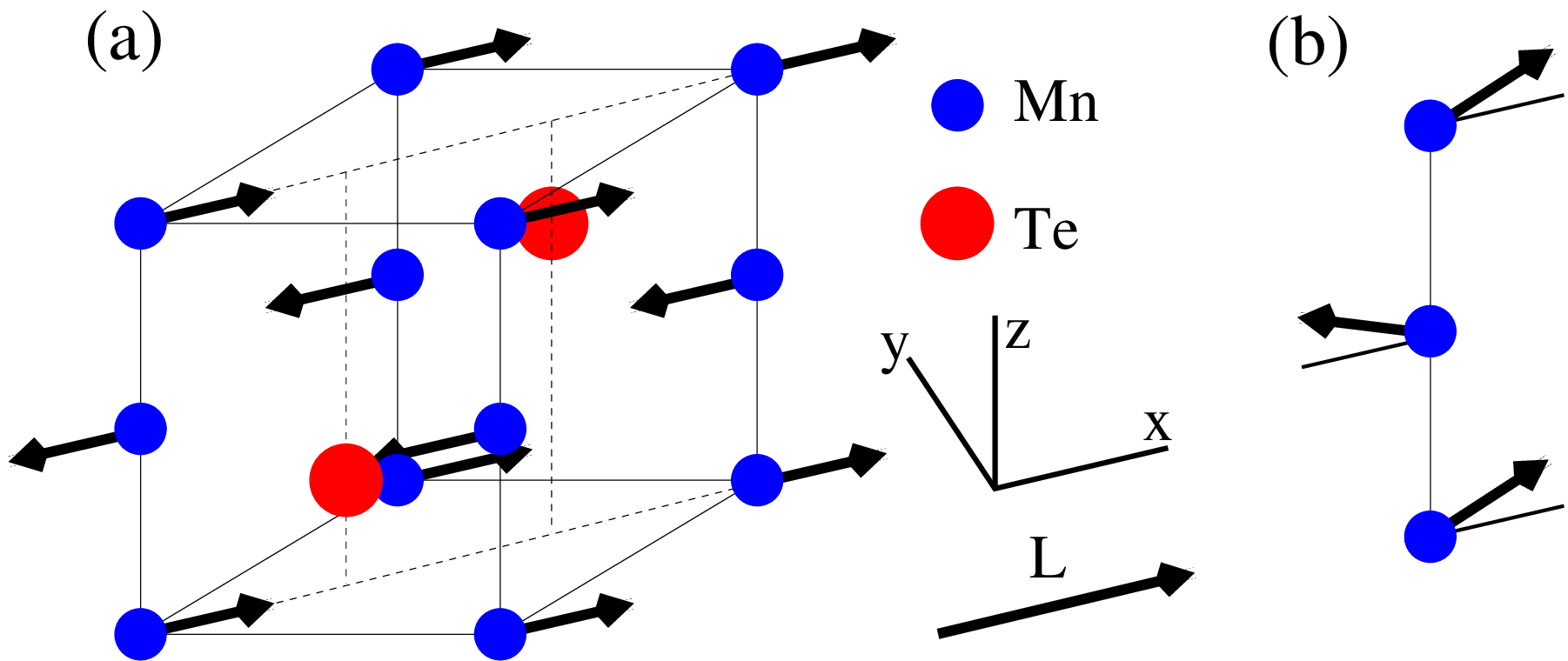}
\caption{(a) Crystallographic and magnetic unit cell of MnTe.
{\bf L} is the Néel vector of the magnetic structure.
In the paper, the $x$ axis is selected parallel to the Néel vector.
(b) Canting of the
Mn atomic moments of the two sublattices leading to weak ferromagnetism.}
\label{fig_magnStr_MnTe}
\end{figure}
The consideration is performed on two different levels.
On the first level, it is carried out in terms of the atomistic magnetic structures,
that is in terms of the directions and values of the atomic moments (Fig.~\ref{fig_magnStr_MnTe}).
Below we
characterize the aspects of the study where the discussion
is carried out on this level.

1. Most of the previous studies of WFM consider spin magnetization only.
On the other
hand, in a number of recent
publications \cite{Jo2025,Sorn2025,Ye2026} the importance of the account for orbital magnetization is
emphasized.
We consider the contributions of both spin and orbital atomic moments.

2. We investigate the role of nonmagnetic ligand atoms.
Besides standard relativistic calculations, we perform calculations where SOC is switched off
either on magnetic metal (Me) atoms or on ligands (Lg).

3. Considerable research attention has been devoted to
the dependence of the WFM characteristics on the strength of SOC \cite{Milivojevic2024,Mazin2024_gossamerWFM,McClarty2024}.
We carry out DFT calculations for various strengths of SOC.
These calculations reveal a complex
nonmonotonous variation of the WFM moment with the SOC strength.

4. We report the results obtained within so-called quasisymemtry (QS) approach.
The term quasisymmetry is used in cases where a part of the interaction is neglected
to obtain the Hamiltonian of a higher symmetry whose analysis provides new physical
insights \cite{Guo2022_quasisym,Li2024_quasisym,Roig2025_quasisym}.
In our case, the QS Hamiltonian contains only one of the three components of the SOC, namely, the component
corresponding to the direction of the Néel vector {\bf L} [Fig.~\ref{fig_magnStr_MnTe}(a)].
The symmetry of the QS Hamiltonian is
intermediate between the symmetries of the nonrelativistic and
relativistic Hamiltonians.
We report a remarkable property of the QS Hamiltonian in the application to AMs:
it does not result in the spin canting and, therefore, the spin WFM is absent.
On the other hand, it gives the origin to the orbital WFM.
This unanticipated result highlights the difference in the physical
mechanisms of the formation of the spin and orbital atomic moments.


To get a deeper insight into the influence of SOC on the properties of AMs
we go on the level of the properties of individual electron states.

5. We introduce the notion
of magnetic structure of electron state (MSES) and study how SOC influences this characteristic
in the case of AM.
We show how the nonrelativistic collinear spin-MSESs of both Me and Lg atoms and nonrelativistic compensated orbital-MSESs
of the Lg atoms obtained in the nonrelativiatic calculation transform
into
complex noncollinear 3D MSESs of both spin and orbital nature, giving large individual contributions to WFM.

6. The analysis of the transformation of the electron states under the influence of SOC
demonstrates the importance of avoided crossings of the electron bands
and suggests an explanation of the
nonmonotonous dependence of the WFM moment on the SOC strength.
The comparison of the LDA and LDA+$U$ calculations shows the influence of Hubbard $U$ on the Me-Lg
hybridization leading to important changes in the properties of MSESs.

Most of the reported calculations
are performed for MnTe, a prototypical material for the class of altermagnets.

\textcolor{black}{
Weak ferromagnetism represents an important phenomenon
for spintronic applications, as it provides a powerful
tool to manipulate antiferromagnetic domains via external magnetic fields \cite{Liu2026}.
The knowledge of the mechanisms of the formation of various contributions to the
WFM moment provides an important support
in the development of new spintronic materials.
}

\textcolor{black}{
The analysis of MSESs of electron states deepens further the understanding of the formation
of relativistic effects in AMs. It shows how the strong magnetic polarization of individual
electron states results in a tiny integral quantity of the WFM moment.
The properties of individual states are important in such experiments as
angle-resolved photoemission spectroscopy (ARPES) restoring electron band
structure. The MSESs provide detailed information for investigation, in its full complexity, the magnetic polarization
of the isoenergy surfaces in the reciprocal space. Such polarization is one of the key
characteristics of the AM materials \cite{Smejkal2002}.
}

\section{Method of calculations}
\label{Sec_Method of calculations}

The calculations are performed with the augmented spherical waves (ASW)
method \cite{Williams1979,Eyert2012}
generalized to deal with noncollinear magnetism and spin-orbit coupling
\cite{Sandratskii1998}.  The local density approximation (LDA) \cite{barth72} and
generalized gradient approximation (GGA)  \cite{Perdew1996} to the
energy functional are employed in the calculations.
The DFT+$U$ method in the form suggested by Dudarev {\it et al.} \cite{Dudarev1998}
was applied to examine the influence of the on-site correlation of the 3d electrons.

\textcolor{black}{
The details of the calculation method are collected in Appendix~\ref{App_sec_Calculation method}.
Besides nonrelativistic and fully relativistic calculations we perform calculations in the quasisymmetry
approach where only one component
of the SOC is taken into account: the component corresponding to the local $z$ axis in Eq.~(\ref{Hso})
that is collinear to the Néel vector of the antiferromagnetic structure.
In most of the calculations, the SOC is taken into account for both Mn and Te atoms.
In some cases, to get an additional physical insight, the SOC is switched off on one
of the atomic types.
}


When the self-consistency is reached, the vectors of spin $\mathbf{m}_{s\mathbf{k}}^\nu$
and orbital
$\mathbf{m}_{o\mathbf{k}}^\nu$
moments
of the $\nu$th atom are calculated for individual electron states as
\begin{equation}
\label{eq_spin_moment_state}
\mathbf{m}_{s\mathbf{k}n}^\nu= \int_{\Omega_{\nu}}\psi_{\mathbf{k}n}^\dag\mathbf{\sigma}
\psi_{\mathbf{k}n}d\mathbf{r}
\end{equation}
\begin{equation}
\label{eq_orbital_moment_state}
\mathbf{m}_{o\mathbf{k}n}^\nu= \int_{\Omega_{\nu}}\psi_{\mathbf{k}n}^\dag\mathbf{\hat{l}}
\psi_{\mathbf{k}n}d\mathbf{r}
\end{equation}
where $\mathbf{\sigma}=(\sigma_x,\sigma_y,\sigma_z)$
and $\mathbf{\hat{l}}=(\hat{l}_x,\hat{l}_y,\hat{l}_z)$, $\psi_{\mathbf{k}n}$ is
the wave function of the Kohn-Sham state corresponding to wave vector $\mathbf{k}$
and band index $n$.
The integrals are carried out over $\nu$th atomic sphere.
We introduce the notions of spin and orbital magnetic structures of electron state (MSESs) given
by the sets of vectors $\{\mathbf{m}_{s\mathbf{k}n}^\nu\}$ and $\{\mathbf{m}_{o\mathbf{k}n}^\nu\}$
for an individual electron state ${\mathbf{k}n}$. The atomic orbital moments are calculated in
atom-centered approximation \cite{Ovesen2024}.


To better understand the influence of the SOC on physics of AMs we perform calculations with various
strengths of SOC. In this case the SOC operator (Eq.~\ref{Hso}) is multiplied by scaling factor $\gamma$
varying in the interval from 1 to 5.

{\it Symmetry analysis.}
The symmetry arguments play an important role in the analysis of the properties of AMs.
Already the definition
of an AM is based on the symmetry-governed property valid in the case that SOC is considered negligible.
Also the relativistic properties such as WFM are governed by symmetry though
the symmetry group of the relativistic
problem is different from the symmetry group characterizing the nonrelativistic case.
The quasisymmetry DFT calculations
can be considered as a physical model intermediate between nonrelativistic and fully relativistic considerations.
The coherent symmetry analysis of all three cases can be performed on the basis of the spin-space
groups. The SSGs allow different transformation of the spin and orbital
variables \cite{Brinkman1966A,Brinkman1966B,Cracknell1974,Chen2024,commentSSG}.
Further information on the symmetry groups relevant for the studies of AMs is given
in Appendix~\ref{App_sec_symmetry}.

{\it Calculation of weak ferromagnetic state.}
 Weak ferromagnetism is the phenomenon of the instability of a collinear antiferromagnetic (AFM) structure with respect to the
    canting of the atomic moments leading to a nonzero magnetization of the material.
    Starting the DFT calculation with a compensated collinear magnetic structure, the questions to
address are the following:
    1.~Can we predict whether
    the initial compensated collinear magnetic structure becomes unstable during iterations?
    2.~If an instability is predicted, is the character of the changes in the magnetic configuration
    a part of the prediction?
The answers to both questions are positive.
\textcolor{black}{
The method of the prediction and its application to AFM with NiAs crystal structure are discussed in Appendix~\ref{App_sec_symWFM}.
In the nonrelativistic case, the collinear spin configuration remains stable and neither spin nor orbital WFM appears.
In the relativistic case, the collinear configuration is unstable in the calculations resulting in canting of both spin and orbital
moments and leading to the appearance of spin and orbital WFMs.
The analysis of the quasisymmetry calculations
shows that the atomic spin moments remain collinear to the {\bf L} axis whereas the
atomic orbital moments deviate from the {\bf L} axis and contribute to the weak ferromagnetism
of the material.
}

\section{Results of calculations}

Most of the results included in the paper are obtained for MnTe with {\bf L}$||$$x$ since this direction of {\bf L}
leads to WFM and corresponds to the experimental situation for MnTe. However, to emphasize a general character
of the conclusions, we include some results obtained for CrSb. The calculations of CrSb were also performed for {\bf L}$||$$x$
since we are interested in the properties of the induced WFM moments.
The rule of symmetry constraint (see Ref.~\onlinecite{Sandratskii2001} and Appendix~\ref{App_sec_symWFM}) allows the
performance of such self-consistent calculation though the experimental ground-state Néel vector of CrSb is collinear to the $z$ axis and WFM
does not appear.

\subsection{Magnetic moments as functions of the SOC strength}
\label{sec_moments_of_gamma}
\begin{figure}[t]
\includegraphics*[width=8cm]{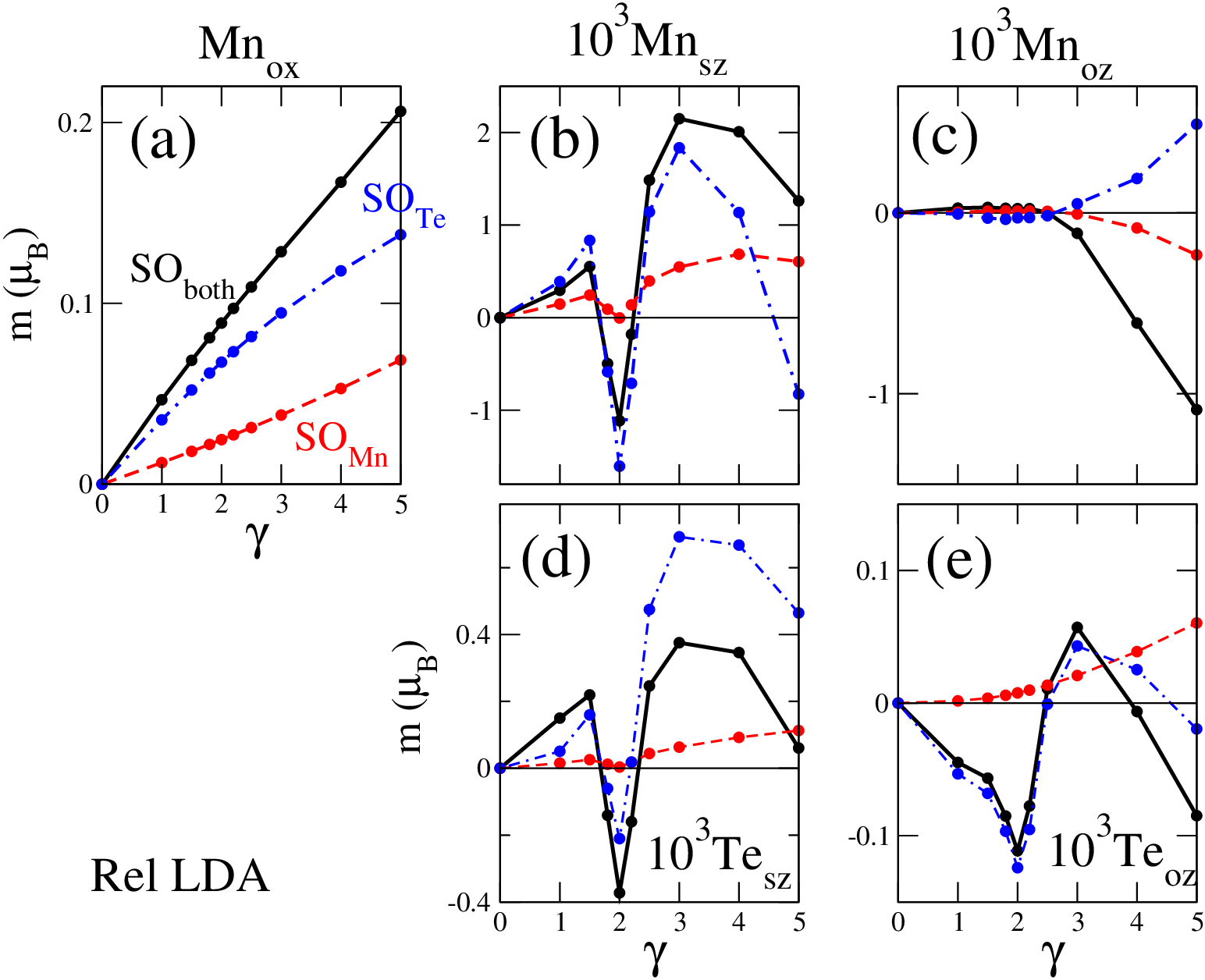}
\caption{Induced moments as functions of the SOC strength. LDA calculation for MnTe. {\bf L}$||$$x$.
(a) The component of the Mn orbital moment collinear to the Néel vector.
(b)-(e) The components of the Mn spin (b) and orbital (c) and Te spin (d) and orbital (e) moments
collinear to the $z$ axis and contributing
to WFM. These components are multiplied by $10^3$. In all graphs the results of the calculations with SOC on
Mn atoms only are shown in red,
with SOC on Te atoms only are in blau, and with SOC on both atoms are in black.
}
\label{fig_mom_of_SOC_MnTe_LDA}
\end{figure}
Figure~\ref{fig_mom_of_SOC_MnTe_LDA} shows the dependencies of the SOC-induced components of the atomic
moments on the SOC scaling parameter $\gamma$ obtained in the LDA calculation for MnTe.
Figures~\ref{fig_mom_of_SOC_MnTe_LDA+U}~and~\ref{fig_mom_of_SOC_CrSb_GGA} of Appendix~\ref{App_sec_moments_of_gamma}
present the same type of dependencies obtained with LDA+$U$ method for MnTe and with GGA method for CrSb.
The calculations show that
there are important qualitative features common for all three cases.

For actual value of SOC, i.e. for $\gamma$=1, the values of the components contributing to the WFM moment
are small: about $\sim$10$^{-3}$ or smaller. This correlates with experimental situation for MnTe.

We begin the discussion of $\gamma$ dependence with the component of the Me orbital moment collinear to the
{\bf L} vector that is collinear to the nonrelativistic direction of the spin moment
[Figs.~\ref{fig_mom_of_SOC_MnTe_LDA}(a),~\ref{fig_mom_of_SOC_MnTe_LDA+U}(a),~\ref{fig_mom_of_SOC_CrSb_GGA}(a)].
The calculations gave rather simple monotonous dependencies that in most
cases are close to a linear function in a broad $\gamma$ interval.
For both compounds,
the absolute values of the Me-SOC and Lg-SOC contributions are comparable.
Hubbard $U$ does not lead to qualitative changes
in the $\gamma$-dependence of the {\bf L} component of the Mn orbital moment
[Fig.~\ref{fig_mom_of_SOC_MnTe_LDA+U}(a)].
More detailed comparison of the results for MnTe and CrSb are given in Appendix~\ref{App_sec_moments_of_gamma}.


Our main interest is in the $z$-components of the moments contributing to WFM.
They are shown in (b)-(e) panels of Figs.~\ref{fig_mom_of_SOC_MnTe_LDA},~\ref{fig_mom_of_SOC_MnTe_LDA+U},~\ref{fig_mom_of_SOC_CrSb_GGA},
There are four contributions: the spin and orbital moments of the Me and Lg atoms.
For each contribution there are three types of functions
obtained with full or partial switching on the atomic SOCs.
In general, different contributions have comparable values,
at least in some $\gamma$ intervals,
and none of the contributions can {\it a priory} be treated as negligible.

%
The analysis of the relativistic calculations (Figs.~\ref{fig_mom_of_SOC_MnTe_LDA},~\ref{fig_mom_of_SOC_MnTe_LDA+U},~\ref{fig_mom_of_SOC_CrSb_GGA})
leads to the following important conclusion.
In contrast to the {\bf L} component of the Me orbital moment [panel (a) of these figures],
the $\gamma$ dependence of the $z$ components is not monotonous. There are very sharp
features in the curves [see panels (b), (d), and (e) of these figures].
Also the signs of the
contributions and their relative signs
change with the variation of $\gamma$.

In QS calculations, the spin WFM does not appear and the number of panels in the corresponding figures
reduces to three (Figs.~\ref{fig_mom_of_SOC_MnTe_LDA+U_QS},~\ref{fig_mom_of_SOC_CrSb_GGA_QS} of Appendix~\ref{sec_QS_moment}).
The comparison of the results of the relativistic and QS calculations
leads to the following two conclusions.
First, the (a) panels of the figures show that the SOC component
collinear to {\bf L} is to a
large extent responsible for the {\bf L} component of the Me orbital moment.
(See Appendix~\ref{sec_QS_moment} for more details).
%
Second,
the sharp nonmonotonous behavior in the $\gamma$-dependencies of the orbital WFM moments
obtained in the relativistic calculations is not obtained in the QS calculations.
This is a signature of the fact that the orthogonal components of the SOC are not only responsible
for the spin WFM but also make induced orthogonal components of the atomic moments
very
sensitive to the value of the SOC. In Sec.~\ref{Sec_Effect_of_SOC_on_ES} we discuss
the origin of this sensitivity and complex behavior.

\subsection{Effect of the SOC on electron states of altermagnets}
\label{Sec_Effect_of_SOC_on_ES}

\textcolor{black}{The small values of the WFM moments for $\gamma$=1 do not
inherently imply a correspondingly small
influence of the SOC on the individual electron states.
}
On the opposite,
the calculations show that the influence of SOC on magnetic structures of electron states, MSESs,
is in general strong, and the electron states
have large components of magnetic moments orthogonal to the Néel vector {\bf L}.

In the consideration of the influence of SOC on individual electron states a principal role
is played by the aspects connected with spin degeneracy of the electron states.
Indeed, according to the principles of quantum mechanics,
the replacement of two degenerate states by two different linear combinations of them provides an
equivalent description of the functional space spanned by the two eigenfunctions.
However, such a replacement changes the MSESs of the two selected basis states.
The situation is principally different in the case of a nondegenerate electron state
where MSES is a uniquely defined and physically significant characteristic.
In this respect there is a principal difference
between AMs and conventional AFMs where the states are Kramers degenerate.

In this paper, we focus on one interval of {\bf k} points [{\bf k}$_-$,{\bf k}$_+$] where {\bf k}$_-$=(0.1,0.2,$-$0.5),
and {\bf k}$_+$=(0.1,0.2,0.5).
Here, the $x$ and $y$ components of the {\bf k} vectors are in units of $\frac{2\pi}{a}$ and the $z$ components are in
units of $\frac{2\pi}{c}$.
With exception of the end and middle points, the interval contains general
{\bf k} points that do not have any spatial symmetry. At such points
the electron states of an AM are spin-split in the nonrelativistic
treatment. The selection of one interval allows, keeping in reasonable limits the amount of presented numerical results,
formulate important findings of the study.

\subsubsection{Comparizon of the band structures obtained in nonrelativistic, quasisymmetry and
relativistic calculations}
\begin{figure}[t]
\includegraphics*[width=8cm]{ 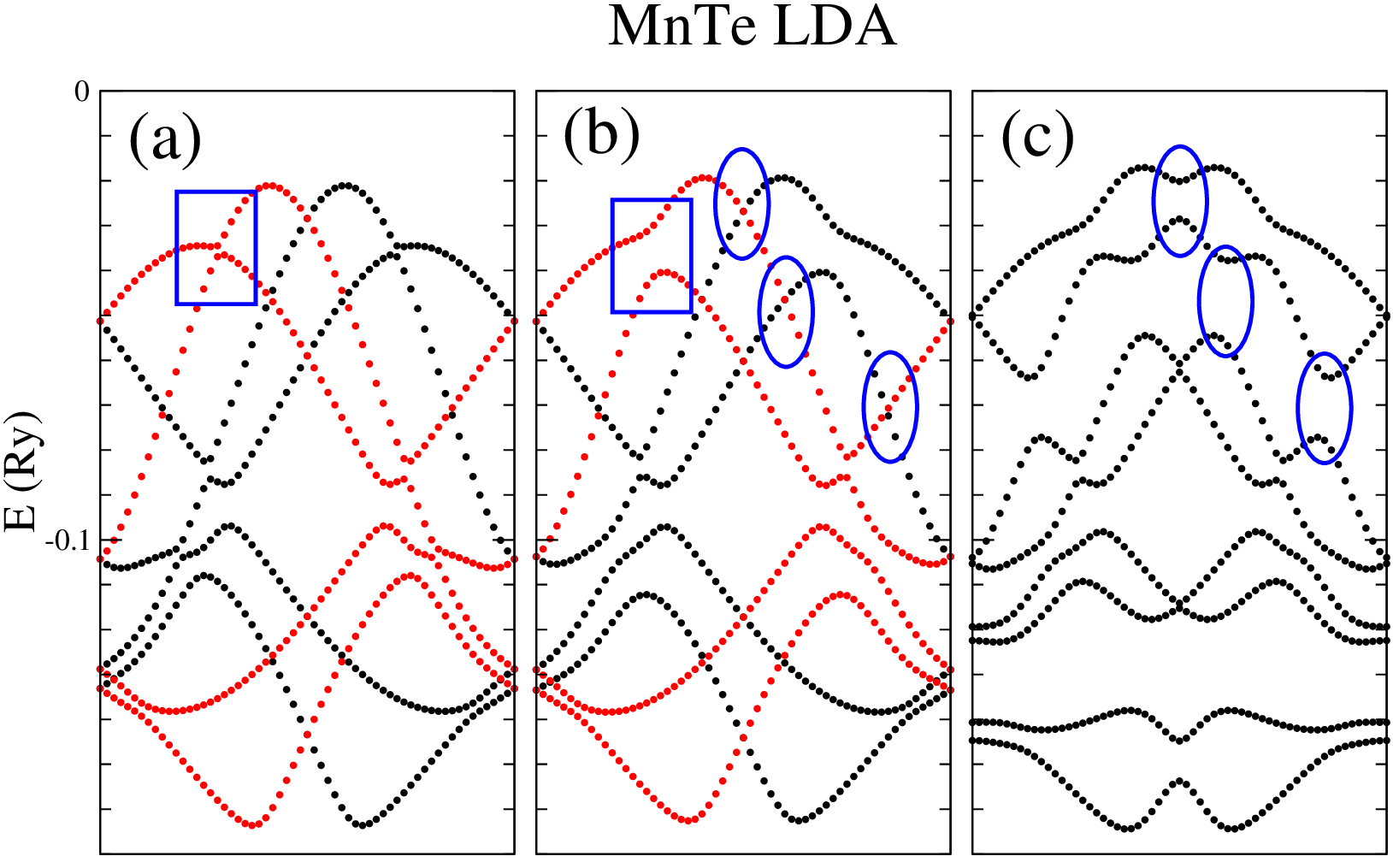}
\caption{Fragment of the band structure of MnTe in the interval [{\bf k}$_-$,{\bf k}$_+$]. LDA calculations.
(a) nonrelativistic, (b) quasisymmetry, (c) relativistic.
Blue boxes highlight some places of avoided crossings of the bands with the same
spin projections. Blue ellipses highlight the places of avoided crossings of bands with opposite spin projections.}
\label{fig_Bands_LDA}
\end{figure}

We start the analysis of the influence of the SOC on the electron states with a general comparison
of the band structures obtained in nonrelativistic, quasisymmetry and
relativistic calculations. Figure~\ref{fig_Bands_LDA} presents the results of the LDA calculations in
the selected {\bf k} interval.
In the NR and QS band structures [Figs.~\ref{fig_Bands_LDA}(a)(b)], the bands have two different colors. In both
these cases the spin projection on the {\bf L} direction is a good quantum number of the electron states, and the
black and red colors
distinguish the spin-up and spin-down bands.
In the relativistic calculation [Fig.~\ref{fig_Bands_LDA}(c)] the states are essentially spin-mixed and are
all presented in the same color.

In the nonrelativistic band structure [Fig.~\ref{fig_Bands_LDA}(a)],
there is no intersections of the bands
of the same color.
This result is expected since we consider
an interval of general {\bf k} points.
The states with the same spin projection correspond to the same irreducible representation (irrep)
of the SSG symmetry group of the {\bf k} point.
This leads to
avoided crossings of the bands of the same color. On the other hand, the states with opposite spin projections
correspond to different irreps of the group of spin rotations about the {\bf L} axis and do not
mix with each other. As a result, the bands with different colors intersect.

The QS calculation [Fig.~\ref{fig_Bands_LDA}(b)] shows that the repulsion of the bands with the same spin projection
at the points of avoided crossings is enhanced by the {\bf L}-component of the SOC. On the
other hand, all intersections of the bands with opposite spin projections remain intact.

Finally, in the relativistic calculation [Fig.~\ref{fig_Bands_LDA}(c)] all states are spin-mixed and belong to the same
trivial irrep of the group containing only unity transformation. The pure spin rotations are not
symmetry operations in this case.
As a result, all crossings of the bands with opposite spin projections are
avoided. The formation of avoided crossings, generated by SOC, is an
important feature that we return to in Sec.~\ref{Sec_MSES}.

The LDA+$U$ calculations presented in Fig.~\ref{fig_Bands_LDA+U} of Appendix~\ref{App_sec_LDA+U} gave qualitatively similar picture of the
relation between SOC and avoided crossings.
However, Hubbard $U$ influences the character of the
Me-Lg hybridization that is an important factor in the properties of the electron states of AMs.
This aspect will be further discussed in Sec.~\ref{Sec_MSES}$b$.

\subsubsection{Magnetic structures of electron states}
\label{Sec_MSES}

In Sec.~\ref{Sec_Method of calculations} we introduced the notion of MSES.
In this section we discuss MSESs of the electron states of MnTe.  First, we present the properties of MSESs
following from the symmetry analysis.
Three different levels of the
treatment of the SOC are considered.

\begin{figure}[t]
\includegraphics*[width=6cm]{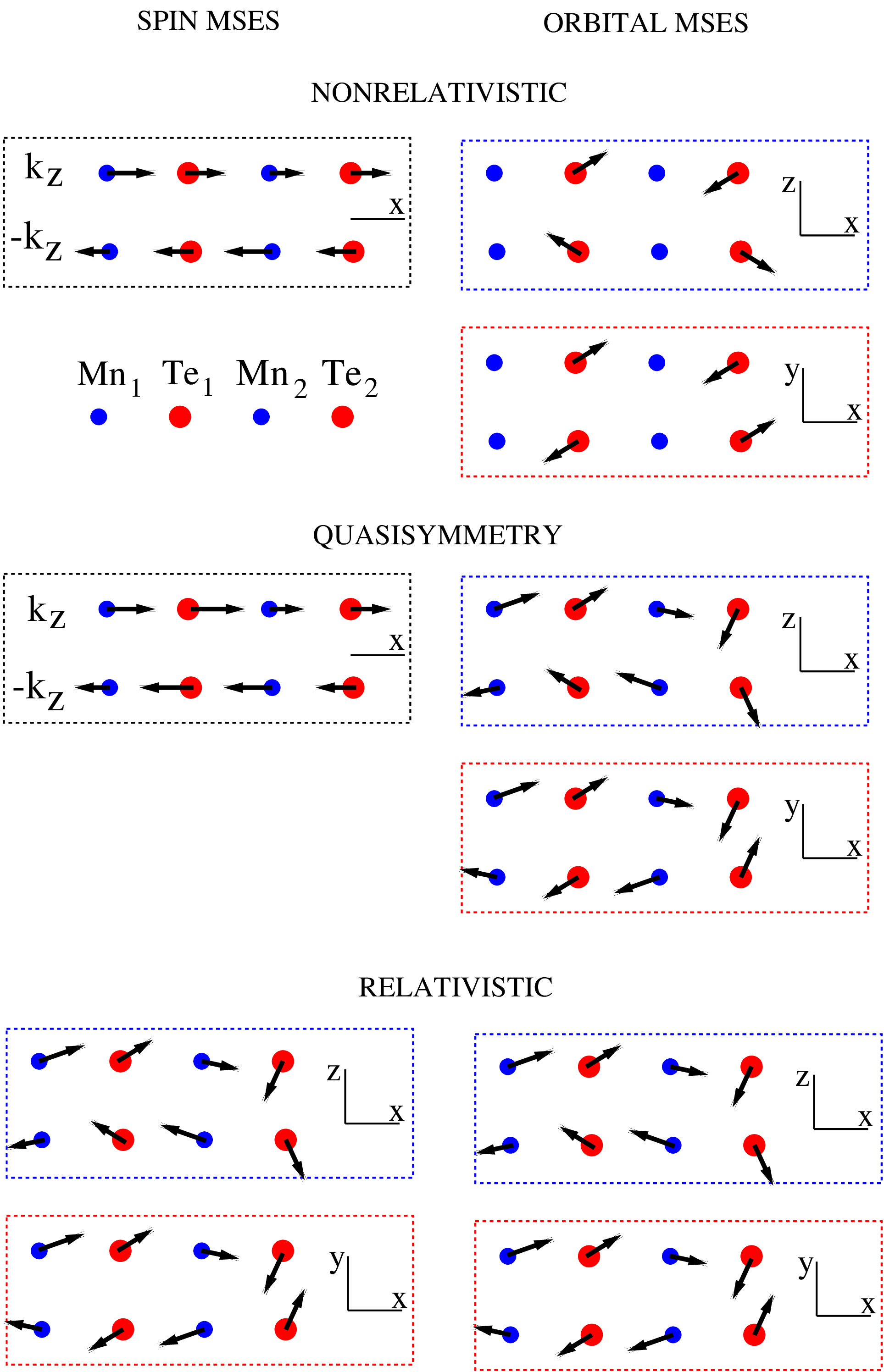}
\caption{Schematic picture of the spin and orbital magnetic structures of electron states
in nonrelativistic, quasisymmetry, and relativistic calculations. Left (Right) column presents spin (orbital) SMESs.
In each rectangle, there are projections of the moments of four atoms for electron states at two {\bf k} points with opposite
values of the $k_z$ component. The black, blue, and red rectangles contain projections on the $x$ axis, $xz$ plane,
and $xy$ plane, respectively.
}
\label{fig_MSES_systematics}
\end{figure}
\paragraph{Symmetry analysis.}
A schematic picture of the variation of MSES
obtained on the basis of the symmetry analysis is presented in Fig.~\ref{fig_MSES_systematics}.
For simplicity,
the four atoms in the unit cell of MnTe are depicted as lying on
one line parallel to the {\bf L} axis, {\bf L}$||$$x$. The atomic positions on the line correspond to the projections
of the atoms on the $xy$ plane.
The arrows present the directions of the atomic moments.

An important role in the symmetry relations between the MSESs at the {\bf k} points of interval [{\bf k}$_-$,{\bf k}$_+$]
is played by the symmetry operation of the
reflection in the $z$=0 plane (operation 8 in Table~\ref{tab_sym}). This operation connects
the {\bf k} points having opposite values of the $z$ projection, $k_z$ and $-$$k_z$.
It transforms the Mn sublattices into each other,
and each Te sublattice into itself.

In the NR case (the upper part of Fig.~\ref{fig_MSES_systematics}), the spin projection on the {\bf L} axis is a good quantum number
of the electron state. Therefore,
the directions of the spin moments of all four atoms in MSES are the same.
Since the system has zero magnetization, the magnetic moments of different states must compensate each other.
In the interval [{\bf k}$_-$,{\bf k}$_+$]
the compensation takes place between two {\bf k} points with opposite $z$ projections.
The symmetry relations are different for Mn
\begin{equation}
{m}_{sx}(k_z,{\rm Mn}_1)=-{m}_{sx}(-k_z,{\rm Mn}_2),
\end{equation}
and Te
\begin{eqnarray}
\label{eq_nr_sym_spin_Te_2}
\begin{array}{rcl}
{m}_{sx}(k_z,{\rm Te}_i)&=&-{m}_{sx}(-k_z,{\rm Te}_i),\:\:\:i=1,2, \\
{m}_{sx}(k_z,{\rm Te}_1)&=& {m}_{sx}(k_z,{\rm Te}_2)\\
\end{array}
\end{eqnarray}
atoms.
They tell us that
the Mn$_1$ and Mn$_2$ moments of the $k_z$ state are compensated by, respectively, the Mn$_2$ and Mn$_1$ moments of
the $-$$k_z$ state. The values of the Mn$_1$ and Mn$_2$ moments of the same state are not connected
by any symmetry restriction and, therefore, are different. For the Te atoms, the spin compensation also takes place between
pairs of states at the {\bf k} points with
opposite $z$ projections but at the same atomic sublattice.
In contrast to Mn moments, the moments of Te$_1$ and Te$_2$ in the same MSES are equal.

The fact that two Te atoms in the NR MSES have equal spin moments is nontrivial since
we deal with general {\bf k} points and there is no nontrivial spatial point transformation
leaving the point {\bf k} invariant and leading to internal symmetry restriction on the MSESs.
The explanation is the following: in the nonrelativistic problem where the complex
conjugation is a symmetry operation of the {\it real} Hamiltonian, the state at any {\bf k} point
is invariant with respect to the combination of the complex conjugation and space inversion.
It is important for altermagnetism of MnTe that the inversion
transforms each Mn sublattice into itself. On the opposite, the Te sublattices are transformed by the inversion into each other.
This results in the equivalence of the Te atoms in the nonrelativistic spin-SMESs at general {\bf k} points.
In the QS case, discussed next, this feature is absent since the {\bf L}-component of SOC contains imaginary unit $i$
and the complex conjugation is not a symmetry operation of the Hamiltonian.

The situation with the orbital atomic moments in NR MSESs is also nontrivial. Although the
nonrelativistic Hamiltonian is real
this does
not immediately mean that for each electron state the atomic orbital moments are zero. The
spin rotations about the {\bf L} axis making the spin projection a good quantum number do not act on the
orbital degrees of freedom.
For the considered {\bf k} interval
the analysis leads to the conclusion
that the orbital moments of the Mn atoms are zero, whereas for the Te orbital moments we obtain
two relations
\begin{equation}
\label{eq_nr_sym_orb_Te_1}
{\bf m}_o(k_z,{\rm Te}_1)=-{\bf m}_o(k_z,{\rm Te}_2)
\end{equation}
and
\begin{eqnarray}
\label{eq_nr_sym_orb_Te_2}
\begin{array}{c}
{m}_{ox}(k_z,{\rm Te}_i)=-{m}_{ox}(-k_z,{\rm Te}_i),\\
{m}_{oy}(k_z,{\rm Te}_i)=-{m}_{oy}(-k_z,{\rm Te}_i),\\
{m}_{oz}(k_z,{\rm Te}_i)= \:\:{m}_{oz}(-k_z,{\rm Te}_i),\\
\end{array}
\end{eqnarray}
where $i$=1,2.
Therefore, the Te atomic orbital moments are 3D vectors with three nonzero Cartesian
components. The orbital moments of the two Te atoms compensate each other within the same electron state.

In the QS calculation, the properties of the spin-MSESs are similar to the properties of the nonrelativistic
spin-MSESs (Fig.~\ref{fig_MSES_systematics}).
\textcolor{black}{
The only difference:
the spin moments of the Te atoms of the same state
become different as the consequence of the loss
of the complex conjugation as a symmetry operation.
}
A qualitatively new feature of the QS electron states,
compared to the NR states, is the presence of orbital weak ferromagnetism. The orbital atomic
moments are 3D vectors with three nonzero Cartesian components. In contrast to the NR case, the orbital moments
are nonzero for both Mn and Te atoms. For the Te moments, only relations given by Eq.~(\ref{eq_nr_sym_orb_Te_2})
remain valid. A similar set of relations
\begin{eqnarray}
\label{eq_nr_sym_orb_Mn_2}
\begin{array}{c}
{m}_{ox}(k_z,{\rm Mn}_1)=-{m}_{ox}(-k_z,{\rm Mn}_2),\\
{m}_{oy}(k_z,{\rm Mn}_1)=-{m}_{oy}(-k_z,{\rm Mn}_2),\\
{m}_{oz}(k_z,{\rm Mn}_1)= \:\:{m}_{oz}(-k_z,{\rm Mn}_2),\\
\end{array}
\end{eqnarray}
is valid for the Mn orbital moments.
There is the compensation of the $x$ and $y$ components
of the orbital atomic moments.
For the $z$ components of the moments there is no symmetry-caused
compensation. These components contribute to the orbital WFM.

In the relativistic case the symmetry-governed structures of both spin-MSESs and orbital-MSESs (Fig.~\ref{fig_MSES_systematics})
are the same as in the case of orbital-MSES of the QS model [Eqs.~(\ref{eq_nr_sym_orb_Mn_2}),(\ref{eq_nr_sym_orb_Te_2})].

\textcolor{black}{The exact symmetry properties of the spin- and orbital-MSES (Fig.~\ref{fig_MSES_systematics})
are an important tool in the studies of the electron states of AMs.
They help to control and analize
the results of first-principles calculations.}

\begin{figure}[t]
\includegraphics*[width=8cm]{ 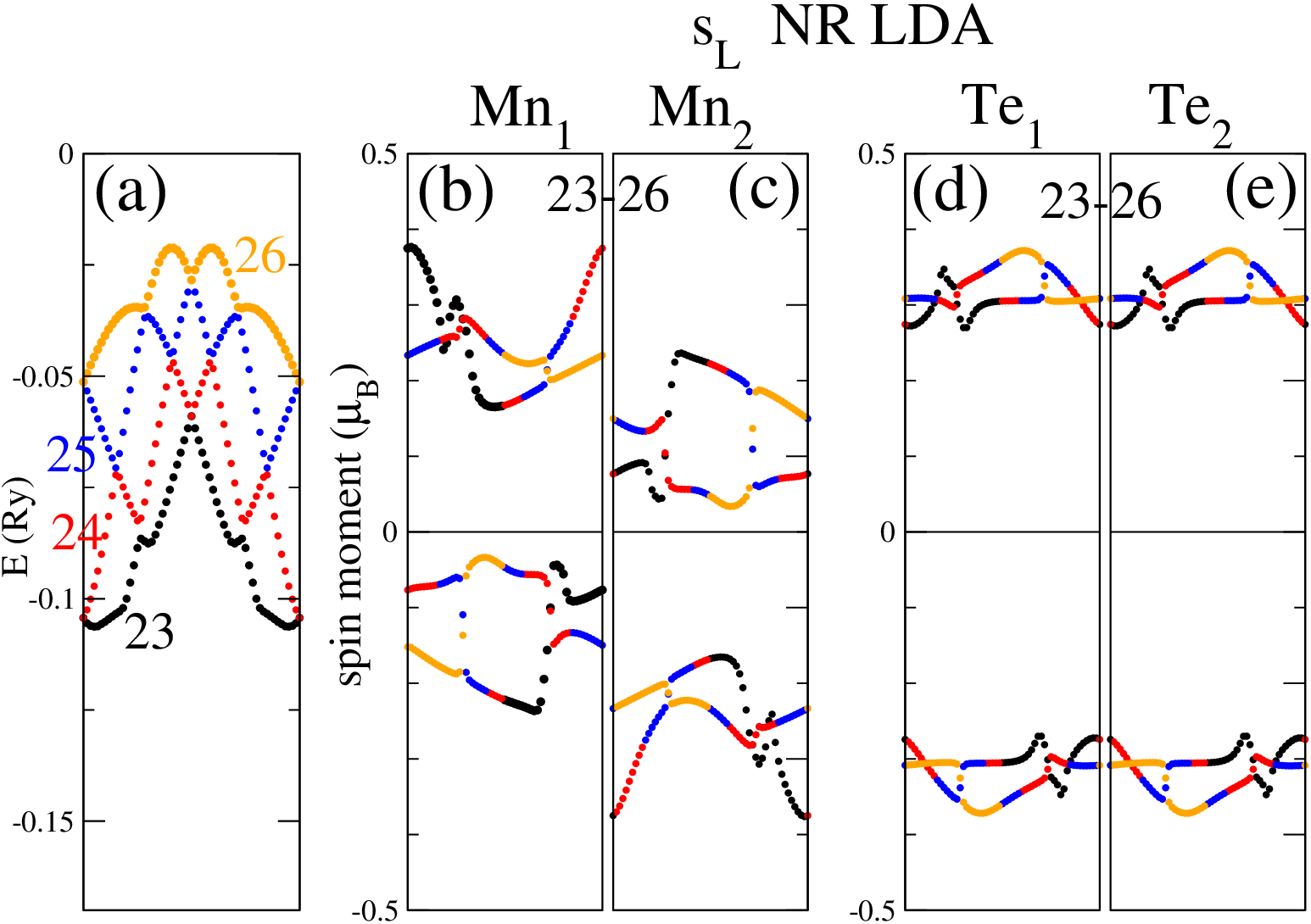}
\caption{Nonrelativistic spin-MSES for bands 23-26 of MnTe calculated in LDA.
(a) Four upper filled bands numbered according to increasing energy of the electron states.
In general numbering of the valence bands, these bands have numbers from 23 to 26.
Different bands have diferent colors.
The abscissa axis of all graphs corresponds to the {\bf k} points from interval [{\bf k}$_-$,{\bf k}$_+$].
(b)-(e) Projections of the atomic moments in the electron states on the axis of the Néel vector
that, in the considered case, is the $x$ axis.}
\label{fig_MSES_SO0_LDA}
\end{figure}

\paragraph{Nonrelativistic calculation of MSESs.}
In Figs.~\ref{fig_MSES_SO0_LDA}~and~\ref{fig_MSES_SO0_LDA+U} we present the spin-MSESs of four upper filled bands
obtained in nonrelativistic LDA and LDA+$U$ calculations.
For convenience of reference, it is useful to number the bands. The conventional method is
to number the electron states at each {\bf k} point in the order of increasing energy.
In Figs.~\ref{fig_MSES_SO0_LDA}~and~\ref{fig_MSES_SO0_LDA+U}, the bands with different numbers are shown using different
colors.
Figures~\ref{fig_Bands_LDA}~and~\ref{fig_Bands_LDA+U}
considered earlier include these bands but with colors corresponding to the
sign of the spin projection. The band obtained by the numbering according to the
energies of the states consists of the pieces of bands with different spin projections.

For the states of four bands, we calculate at each {\bf k} the MSESs and plot
the corresponding values of four atomic moments as shown in Figs.~\ref{fig_MSES_SO0_LDA}(b)-(e)
using the colors of the corresponding numbered bands.
As a result, we obtain the {\bf k} dependence of MSESs of two spin-up and two spin-down bands composed of the
pieces corresponding to the bands with different numbers.

\textcolor{black}{
In the collinear nonrelativistic calculation the value of the atomic spin moment
in a given MSES
corresponds to the partial contribution of the atom into the electron state. The sum
of the contributions of four atoms in the MSES is exactly unity [Figs.~\ref{fig_MSES_SO0_LDA}(b)-(e)].
Therefore, the values of the spin
moments of different atoms characterize the interatomic hybridization in the given electron state.
}

It is instructive to compare the spin-MSESs obtained in the LDA (Fig.~\ref{fig_MSES_SO0_LDA})
and LDA+$U$ (Fig.~\ref{fig_MSES_SO0_LDA+U})
calculations. An important difference between the results of
these two calculations is a strong
decrease of the Mn contribution to the electron states obtained with the LDA+$U$ method.
This effect of Hubbard $U$ is
expected since due to the influence of $U$ the occupied Mn 3d states are
shifted to lower energies.

   The Me-Lg hybridization is important for altermagnetism.
A hypothetical material consisting of only the magnetic Mn sublattices of MnTe would
be a conventional AFM.
In both LDA and LDA+$U$ calculations the hybridization of the atomic Mn and Te states
in the crystal states shown in Figs.~\ref{fig_MSES_SO0_LDA}~and~\ref{fig_MSES_SO0_LDA+U}
is strong. Correspondingly
strong is the nonrelativistic altermagnetic spin-splitting (Figs.~\ref{fig_Bands_LDA}~and~\ref{fig_Bands_LDA+U}).
Remarkably, in both cases a larger part of
the spin moment of the states comes from the Te atoms. In the LDA+$U$ calculation, this latter feature is especially
strongly pronounced.

Figure~\ref{fig_MSES_SO0_LDA+U} shows the calculated three Cartesian components of the orbital moments of the Te
atoms. As predicted by symmetry analysis, the Te moments are the 3D vectors satisfying the symmetry relations given
by Eqs.~(\ref{eq_nr_sym_orb_Te_1}),(\ref{eq_nr_sym_orb_Te_2}) and presented schematically in Fig.~\ref{fig_MSES_systematics}.

\begin{figure}[t]
\includegraphics*[width=8cm]{ 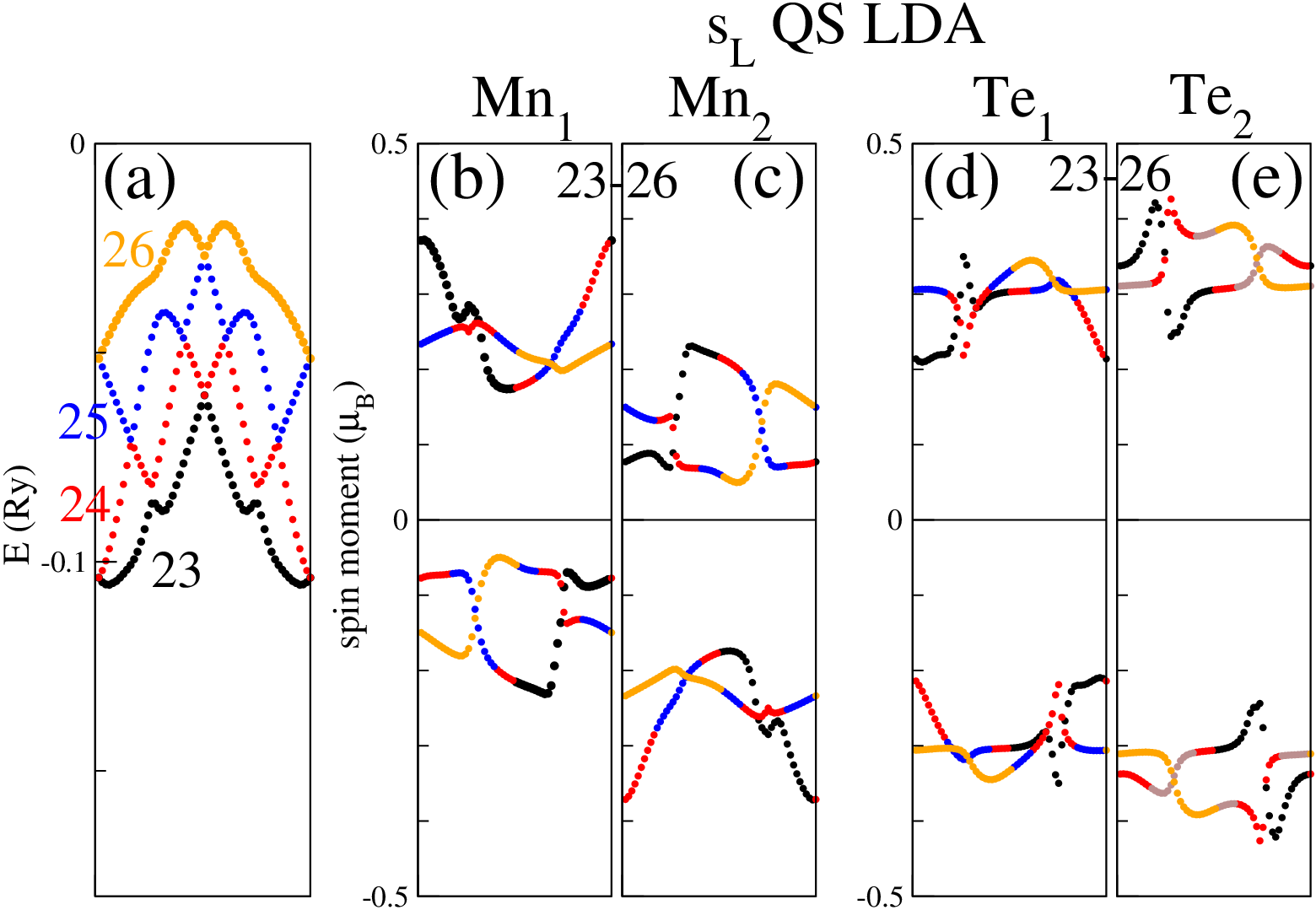}
\caption{The same as in Fig.~\ref{fig_MSES_SO0_LDA} but calculated with quasisymmetry LDA.}
\label{fig_spin-MSES_SO1X_LDA}
\end{figure}

\paragraph{Quasisymmetry calculation.}
As predicted by the symmetry analysis (Fig.~\ref{fig_MSES_systematics})
the properties of
the spin-MSESs obtained in the QS calculations
are qualitatively similar to
the properties of the spin-MSESs of the nonrelativistic calculations
(compare Figs.~\ref{fig_MSES_SO0_LDA}~and~\ref{fig_spin-MSES_SO1X_LDA} for the LDA
calculations and Figs.~\ref{fig_spin-MSES_SO1X_LDA+U}~and~\ref{fig_MSES_SO0_LDA+U}
for the LDA+$U$ calculations).
Thus, the MSESs of two spin-up and two spin-down bands are
composed from the pieces of the bands with different numbers. In contrast to NR calculation,
the Te$_1$ and Te$_2$ atoms become inequivalent.

A qualitatively new feature of the QS calculation compared to the NR calculation
is nonzero orbital atomic moments on both Mn and Te atoms and weak orbital ferromagnetism.
In Figs.~\ref{fig_orbital-Mn-MSES_SO1X_LDA+U}~and~\ref{fig_orbital-Te-MSES_SO1X_LDA+U}
we present the components of the orbital-MSESs.
The $k_z$ dependencies are nonmonotonous and, in contrast to the spin-MSES, change the sign with variation
of $k_z$. The curves of spin- and orbital-MSESs are distinctly different.
However, one can notice
a correlation between specific features in the $k_z$ dependencies of the spin and orbital MSESs with the positions of the enhanced
avoided crossings between the bands with the same spin projection in the QS band structure.

The $z$-components of the orbital moments of the individual MSESs are much larger than the
integrated WFM moments in the atomistic description. This is the consequence of the mutual compensation
of the moments of different states. The presence of the compensation is
well seen in Figs.~\ref{fig_orbital-Mn-MSES_SO1X_LDA+U}~and~\ref{fig_orbital-Te-MSES_SO1X_LDA+U}.
However, for the $z$-component this compensation is never complete since there is no symmetry
operation responsible for a zero value of the $z$ projection of the integrated atomic moment.

\begin{figure}[t]
\includegraphics*[width=8cm]{ 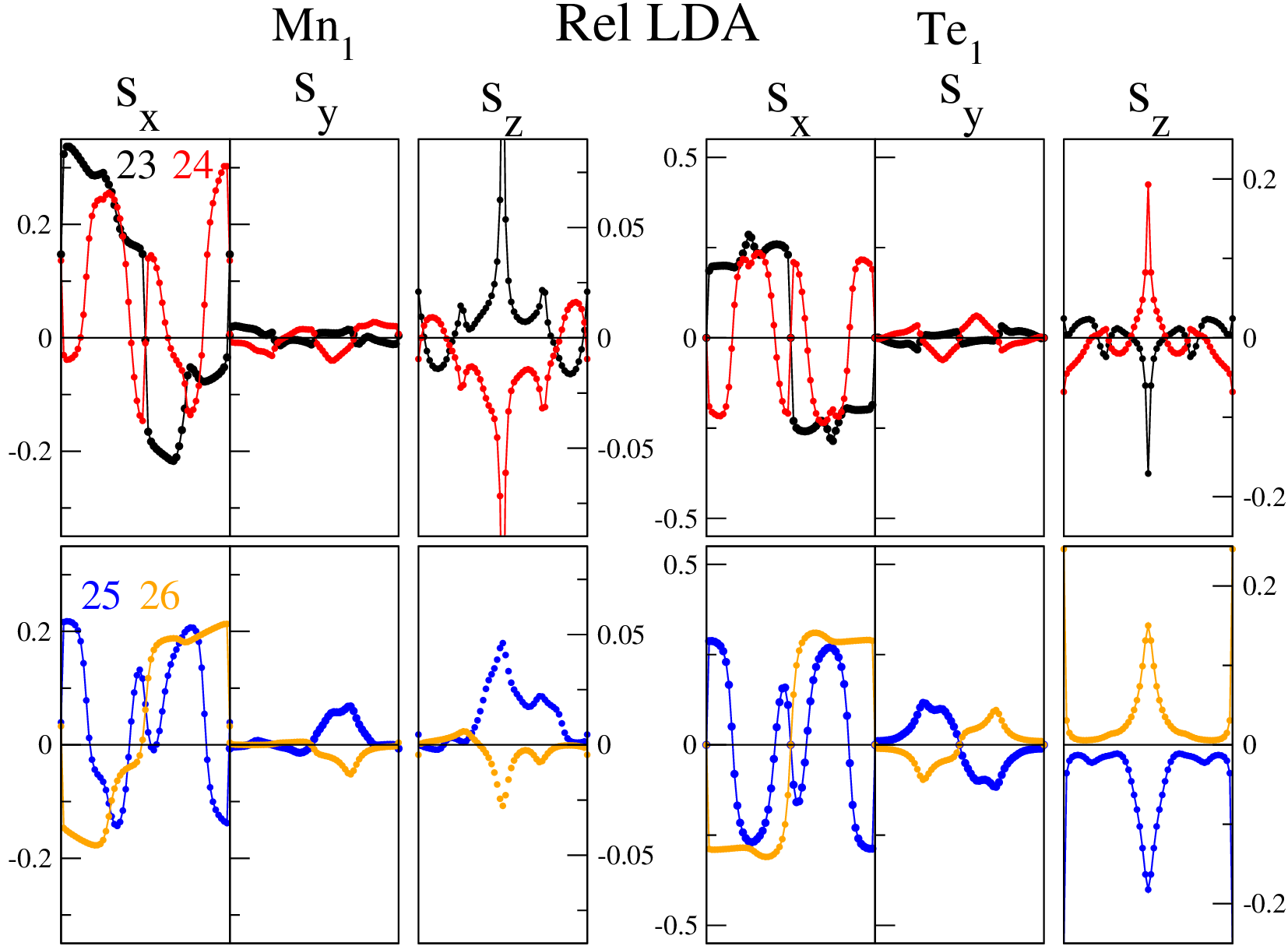}
\caption{Relativistic spin-MSES calculated in LDA.
Shown are three Cartesian components of the atomic spin moments.
All ordinate axes give the value of the moment in units of magneton Bohr ($\mu_{\rm B}$). Left (Right)
part presents the Mn$_1$ (Te$_1$) atom. The upper 6 panels are for the electron bands with
numbers 23 and 24, the lower 6 panels are for the bands with numbers 25 and 26.
}
\label{fig_spin-MSES_SO1_LDA}
\end{figure}
\begin{figure}[t]
\includegraphics*[width=8cm]{ 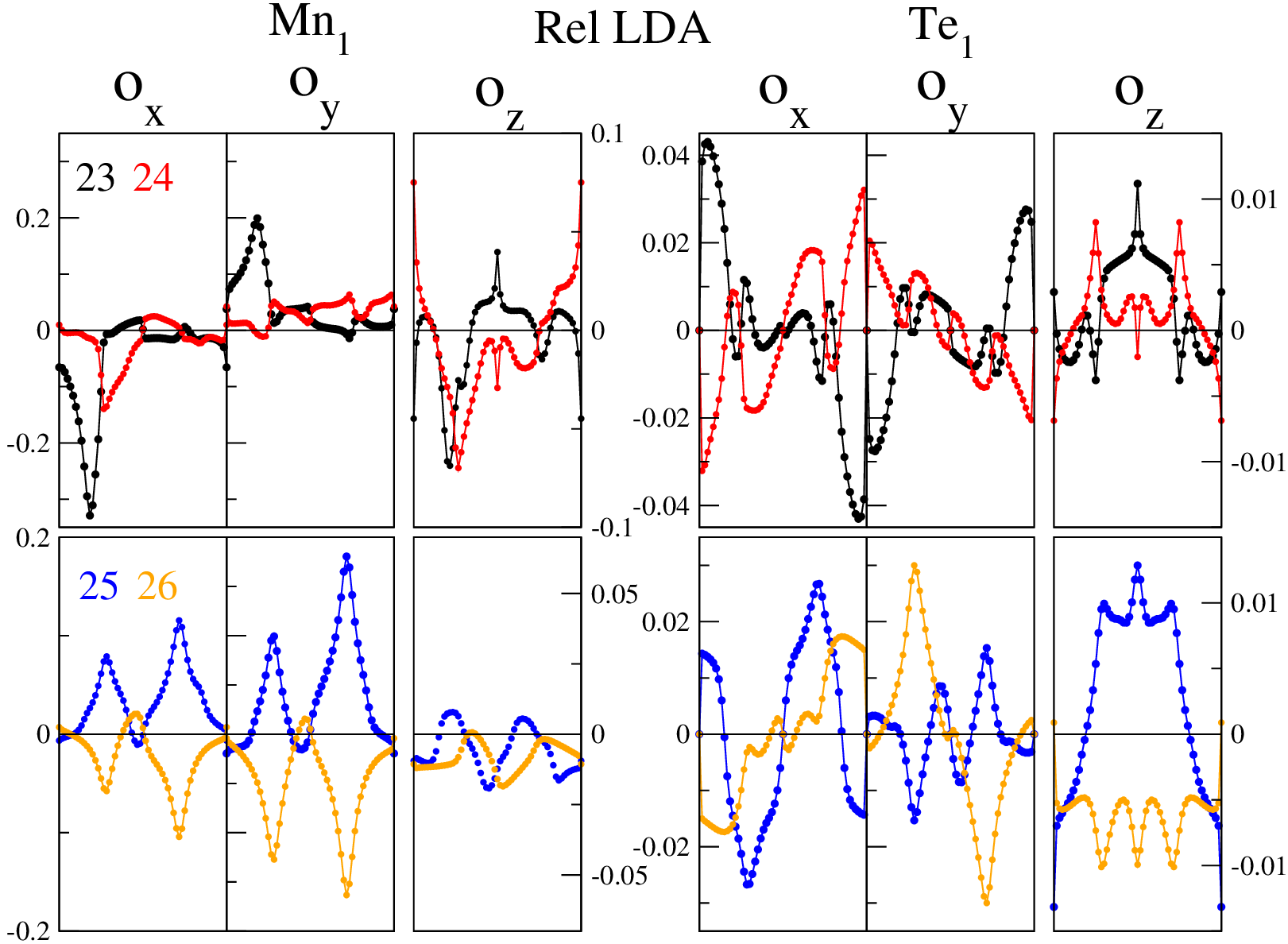}
\caption{The same as in Fig.~\ref{fig_spin-MSES_SO1_LDA} but showing orbital-MSES
instead of spin-MSES.}
\label{fig_orbital-MSES_SO1_LDA}
\end{figure}

\paragraph{Relativistic calculation.}
The symmetry properties of the relativistic spin and orbital MSESs
are identical to the symmetry properties of the orbital MSESs of the QS approach.
Figures~\ref{fig_spin-MSES_SO1_LDA}~and~\ref{fig_spin-MSES_SO1_LDA+U} show the spin-MSESs
for the LDA and LDA+$U$ calculations. To restrict
the number of
graphs, only the contributions of Mn$_1$ and Te$_1$
are presented.
As accidental band intersections are avoided in the relativistic calculations,
all continuous MSES curves have a certain color corresponding to one of the numbered bands.
The corresponding energy bands are presented in Figs.~\ref{fig_Bands_LDA}(c)~and~\ref{fig_Bands_LDA+U}(c).

In drastic difference to the NR and QS spin-MSESs
here the spin-MSESs are complex 3D magnetic structures
(Fig.~\ref{fig_MSES_systematics}).
The $k_z$ dependencies of the spin components are, in general,
also very complex (Figs.~\ref{fig_spin-MSES_SO1_LDA}~and~\ref{fig_spin-MSES_SO1_LDA+U}).
All components
change sign several times in the considered {\bf k} interval.
This demonstrates that, in contrast to the NR and QS calculations, we deal here with the
bands of strongly spin-mixed states.
Especially strong changes of the curves are in the $k_z$-regions
where in the relativistic band structure the accidental band intersections are replaced by avoided crossings.
Again, the $z$ components of the atomic moments in the MSESs of
individual states are large compared to the small integral values of the WFM atomic moments, and
there is a clear trend to mutual
compensation of the $z$ components of the MSES of different states.
The compensation is never complete.

The sign of the integrated WFM moment cannot be predicted on the symmetry basis and needs direct calculation.
The following consideration is helpful for understanding of the formation of the spin components orthogonal
to the Néel vector. The {\bf L} axis is chosen as a spin-quantization axis. The orthogonal SOC components taken into
account in the relativistic calculation lead to the mixing of the spin-up and spin-down states.
Let us consider a spin-mixed spinor
$\left(\begin{array}{c}\psi_1\\\psi_2e^{i\phi} \end{array}\right)$.
The orthogonal spin component will have the form
$\psi_1\psi_2\sin\phi$. Here $\psi_1$ and $\psi_2$ are nonzero real numbers and $\phi$ is a phase factor.
The sign of the orthogonal spin component depends on the phase factor $\phi$ of the spin mixing.
If $\psi_1^2+\psi_2^2=1$,
the value of the orthogonal component is maximal for $\psi_1=\psi_2$, that is for the strongest spin mixing
leading to zero spin projection on the {\bf L} axis.
The phase factors $\phi$
is the result of the complex hybridization processes caused by SOC and leading to the modification of
the crystal electron states.
The spin-mixing is very strong in the {\bf k}-regions of avoided crossings.
In Figs.~\ref{fig_spin-MSES_SO1_LDA}~and~\ref{fig_spin-MSES_SO1_LDA+U}
the strong influence of the avoided crossings on the
spin projections of the calculated electron states is clearly reflected.

Also in the properties of the components of the orbital moments
shown in Figs.~\ref{fig_orbital-MSES_SO1_LDA}~and~\ref{fig_orbital-MSES_SO1_LDA+U} we obtain complex behavior
correlating with the positions of avoided crossings.
However, the formation of the atomic orbital moments is principally different from the formation
of the atomic spin moments. We can see this on the basis of Eq.~(\ref{eq_orbital_moment_state}).
The angular dependence of state $\psi_{\mathbf{k}n}$ in the $\nu$th atomic sphere can be presented
in the form $\sum_m c_{m}Y_{lm}$ where $Y_{lm}$ are complex spherical harmonics. We consider the
contribution corresponding to a given $l$. The values of the atomic orbital moments will have
the form
\begin{equation}
\begin{array}{l}
l_z=\sum_{m=1}^lm(|c_{m}|^2-|c_{-m}|^2),\\\\
l_x=\sum_{m=-l}^{l-1}\Re(c_{m} c_{m+1}^\ast)[(l-m)(l+m+1)]^\frac{1}{2},\\\\
l_y=\sum_{m=-l}^{l-1}\Im(c_{m} c_{m+1}^\ast)[(l-m)(l+m+1)]^\frac{1}{2}.
\end{array}
\end{equation}
Here the $z$ axis is the axis of the quantization of the spherical harmonics that is convenient to
select collinear to the Néel vector {\bf L}. These expressions show that the orbital moment depends on the
coefficients \{$c_{m}$\} of the decomposition of the wave function in the spherical harmonics. In particular,
the $z$ component of the moment depends on the $\pm m$ polarization of the function, that is on the different
occupation of the atomic states with opposite $m$ values. This is very different from the spin-polarization
leading to the spin moment of the state. The hybridization caused by the SOC leads to the modification of the
coefficients \{$c_{m}$\} resulting in the complex behavior of the components of the atomic orbital moments
reflected in Figs.~\ref{fig_orbital-MSES_SO1_LDA}~and~\ref{fig_orbital-MSES_SO1_LDA+U}.

The importance of avoided crossings in the relativistic band structure suggests an explanation of
the nonmonotonous dependence on the SOC strength of the integral value of the weak
ferromagnetic moment obtained in the calculations presented in Sec.~\ref{sec_moments_of_gamma}.
The mutual compensation of the $z$ projections of the atomic moments of different electron states
leads to small integral values of the WFM moment. The change of the SOC strength influences the mixing
of states in all regions of avoided crossings. Such regions are numerous since in the volume of the Brillouin zone (BZ) the
NR electron bands are spin split and intersect.
As the electron states involved in the
formation of avoided crossings are different in different {\bf k} regions,
the dependencies of the WFM contributions
of the regions on the SOC strength are distinct.
The sensitivity to the SOC strength of the relative values of the WFM contributions
coming from different regions of avoided crossings is the source of
the complex nonmonotonous dependence of the integral value of the weak
ferromagnetic moment obtained in the calculations.

\textcolor{black}{
The principles of the SOC influence on the electron states are general and valid for various materials. However, their concrete realization depends
on the system since the position of the accidental crossings in the band structure and the strength of
the relativistic repulsion are material dependent.
The complex {\bf k}-dependence of the atomic contributions to MSES
reflexes potential sensitivity
of the transport electron properties to the doping.
}

\section{Conclusions}
Recently,
the special properties of altermagnets caused by relativistic spin-orbit coupling
became the topic of extensive investigations.
The focus of the paper is on the phenomena of
weak ferromagnetism and relativistic splitting of the electron states.
Three different levels of the account for SOC are considered: nonrelativistic, quasisymmetry and fully
relativistic.
%

The consideration of the magnetic properties of AMs is performed on two different levels.
On the first level, the atomistic magnetic structure of weak ferromagnetic state is calculated.
It is given by the values and
directions of the atomic moments. The calculations show the importance of taking
into account both spin and orbital atomic moments of both magnetic and ligand atoms.
We separate the contributions of the SOC coming from the metal and ligand atoms to WFM
by switching off the SOC on one of the atomic types. Also here the calculations show the
importance of both contributions.

We study the dependence of the WFM moment on the strength of the SOC and obtain a peculiar nonmonotonous
type of dependence.

An unexpected result was obtained in the quasisymmetry calculation:
the conventional spin WFM was proved to be absent whereas the orbital WFM is present.
This result reveals a principal difference
in the formation of the spin and orbital magnetic moments.

On the second level, the study is focused on the properties of individual
electron states caused by SOC that provides deeper insight in the formation of the
relativistic properties of AMs.
We introduce the notion of the magnetic structure of individual electron states (MSES).
It is shown how the collinear spin MSESs and compensated orbital MSESs of the Lg atoms
obtained in the nonrelativiatic calculation
transform in the QS calculation into a combination of collinear spin MSES and complex noncollinear 3D
orbital MSES responsible for orbital WFM to, finally, become the combination of complex 3D spin and complex 3D orbital MSESs in the
relativistic case.
Both Me and Lg atoms in the relativistic MSESs of individual electron states give large contributions to WFM.
These large contributions of different states strongly
compensate each other in the integral values of the spin and orbital WFM moments.

The importance of the Me-Lg hybridization in the formation of the AM
properties is investigated by means of the comparison of the LDA and LDA+$U$ calculations:
Hubbard $U$ shifts the occupied Me 3d states to lower energies changing
the Me-Lg hybridization of individual electron states and considerably modifying MSES of these states.

An important channel of the influence of SOC on the electron states of the AMs is the relativistic
spin splitting of the accidental spin degeneracies present in the nonrelativistic and quasisymmetric band structures.
The splitting leads to the formation
of strongly developed regions of avoided crossings
that are related to the nonmonotonous dependence
of the WFM moment as the function of the SOC strength obtained at the atomistic level of the study.

An important role in all parts of the paper is played by the symmetry analysis on the basis of
the spin space groups.
The application of the SSGs allows obtaining a coherent description of the three types
of calculations: nonrelativistic,
QS and relativistic. From the viewpoint of the SSGs, in this series of calculations,  the symmetry group of
the next case is a subgroup of the symmetry group of the previous one.

\section{Acknowledgements}
L.S. and K.C. acknowledges financial support by P JAK, Ministry of Education, Youth and Sports of the Czech Republic,
project Quantum materials for applications in sustainable technologies (QM4ST), funded as Project No. CZ.02.01.01/00/22\_008/0004572.
K.C. acknowledges financial support of Czech Science Foundation, Grant No. 26-23625S.
V.M.S acknowledges financial support by Grant PID2022-139230NB-I00
funded by MCIN/AEI/10.13039/501100011033.

\appendix
\renewcommand{\thetable}{A\arabic{table}}
\setcounter{table}{0}

\section{Calculation method}
\label{App_sec_Calculation method}

The nonrelativistic Kohn-Sham Hamiltonian adopted for the noncollinear magnetic
configurations has the form
\begin{equation}
\label{eq_KS}
{\bf \hat{H}}_{KS}={\bf \hat{T}}_{kin}+{\bf \hat{V}}
\end{equation}
where ${\bf \hat{T}}_{kin}$ is spin-diagonal operator of the kinetic
energy and ${\bf \hat{V}}$ is the potential which
in the $i$th atomic sphere takes the form
 \begin{eqnarray}
 \label{pot}
{\bf V}_{i}(r)=
 {\bf U}(\theta_i,\phi_i)^\dagger\;
 \left(\begin{array}{cc}V_i^{+}(r) &0\\
 0&V_i^{-}(r)\end{array}\right)
 {\bf U}(\theta_i,\phi_i)
 \end{eqnarray}
where angles $\theta_i$ and $\phi_i$ determine the local spin-coordinate system of the
$i$th atom characterized by the diagonal form of the potential matrix; $V_i^{+}$
and $V_i^{-}$ are the spin-up and spin-down potentials in the local system;
${\bf U}(\theta_i,\phi_i)$ is the matrix of spin-$\frac{1}{2}$ rotation transforming the potential
from the atomic system of the $i$th atom to the global system.

The operator of the spin-orbit coupling is taken in the form \cite{Sandratskii21}
\begin{eqnarray}\nonumber
\label{Hso}
{\bf H}_{so}=\frac{1}{(2c)^2}\frac{1}{r}
     \bigg [
     \left(\begin{array}{cc}
     \frac{1}{M_{+}^2}\frac{dV^{+}}{dr} & 0\\
      0&\frac{1}{M_{-}^2}
      \frac{dV^{-}}{dr}
       \end{array}\right)
\sigma_z\hat{l}_z \\
 + \frac{1}{M_{av}^2}\frac{dV^{av}}
{dr}
   (\sigma_x\hat{l}_x+\sigma_y\hat{l}_y)
     \bigg ]
\end{eqnarray}
where $V^{+}$ and $V^{-}$ are spin-up and spin-down electron potentials,
 \begin{equation}
V^{av}=\frac{1}{2}
(V^{+}+V^{-})
    \label{so2}
 \end{equation}
and
 \begin{equation}
M_\alpha=\frac{1}{2}(1-\frac{1}{c^2}V^\alpha)\quad ,\alpha=av
,+,-.
 \label{so3}
 \end{equation}
$\sigma_x,\sigma_y,\sigma_z$ are the Pauli matrices and $\hat{l}_x,\hat{l}_y,\hat{l}_z$ are
the operators of the components of the orbital momentum,
$r$ is the distance from the center of atomic surface, $c$ is
the light velocity. In Eq.~(\ref{Hso}), it is assumed that the local atomic $z$ axis is parallel to
the direction of the atomic spin moment.

At the end of each iteration, the two-by-two atomic density matrices
\begin{equation}
\label{eq_density_matrix}
\mathbf{\rho}^\nu=\sum_{\mathbf{k}n}^{occ}\int_{\Omega_{\nu}}\psi_{\mathbf{k}n}
\psi_{\mathbf{k}n}^\dag d\mathbf{r}
\end{equation}
are calculated. If the off-diagonal elements of the matrix are nonzero, the new directions
of the spin moments are determined by the diagonalization of the matrix.

Equations~(\ref{eq_spin_moment_state}),(\ref{eq_orbital_moment_state}) of the main text give the definition
of atomic moments of individual
electron states contributing to MSES of this state.
The sum of the moments over occupied states
\begin{equation}
\label{eq_spin_moment}
\mathbf{m}_s^\nu=\sum_{\mathbf{k}n}^{occ}\mathbf{m}_{s\mathbf{k}n}^\nu,
\end{equation}
\begin{equation}
\label{eq_orbital_moment}
\mathbf{m}_o^\nu=\sum_{\mathbf{k}n}^{occ}\mathbf{m}_{o\mathbf{k}n}^\nu
\end{equation}
gives the values of atomic moments in the atomistic
description of the magnetic structure.

\textcolor{black}{
Our choice of the $U$ value for the Mn 3d electrons in the LDA+$U$ calculations was based on the values used
in the previous calculations \cite{Krause2013,Rooj2023,Mazin2024_gossamerWFM,Carmine2025}
where the effective $U$ value varied in the interval
from 2 eV to 4 eV. We performed systematic calculations for $U$=0.2~Ry (2.72~eV) and some calculations for $U$=0.3~Ry (4.08~eV).
We show the numerical results for $U$=0.2~Ry and briefly discuss general trends in the influence of Hubbard $U$ on the electron
states. The symmetry based conclusions are valid for both DFT and DFT+$U$ calculations.
}

\section{Symmetry groups relevant for the studies of AMs}
\label{App_sec_symmetry}

This appendix gives a brief introduction of the SSGs as an adequate tool for
the combined analysis of the nonrelativistic, quasirelativistic and relativistic DFT-based
calculations.
Essential for the application
of the SSGs to the DFT calculations is the consistent treatment of the electron states as spinor functions.
Traditionally, in the nonrelativistic collinear DFT calculations,
the electron states have been considered as scalar
functions supplied with a spin index \cite{Turek2022,Sandratskii1981,Sandratskii2025}.
However, the combined consideration of the nonrelativistic and relativistic
DFT calculations becomes in this case rather cumbersome.

A more detailed introduction of the SSG operators can
be found in Ref.~\onlinecite{Sandratskii2025}.
The element of SSG is written as $\{\alpha_S|\alpha_R|\boldsymbol{\tau}\}$
where $\alpha_S$ and $\alpha_R$ are spin and space rotations respectively, $\boldsymbol{\tau}$ is
space translation.
We introduce notation $[\alpha_R|\boldsymbol{\tau}]$$\equiv$$\{E_S|\alpha_R|\boldsymbol{\tau}\}$
for the orbital part of the operation. Here $E_S$ is the unity spin transformation.

In the nonrelativiatic case the SSG of the Kohn-Sham equation of the AM
has the form of the following direct product
\begin{equation}
G_{NR}={\bf C}_L\times{\bf K}\times G_{tr}\times G_{sub}
\end{equation}
where ${\bf C}_L$ is the group of the pure spin rotations $\{C_{L\phi}|E_R|0\}$
about the {\bf L} axis; group ${\bf K}$ consists of unity transformation $E$ and complex
conjugation $K$; group $G_{tr}$ consists of $E$ and operation
$\left(\begin{array}{cc} 0&-1\\1 &0 \end{array}\right)g_{tr}$
where $g_{tr}=[\alpha_R|\boldsymbol{\tau}_\alpha]$
transforms
magnetic sublattices into each other and matrix $\left(\begin{array}{cc} 0&-1\\1 &0 \end{array}\right)$
reverses the spin projection of the two-component spinor.
To recall, the time-reversal operation in the relativistic consideration has the form of the product \cite{Sandratskii2025}
\begin{equation}
\Theta=\left(\begin{array}{cc} 0&-1\\1 &0 \end{array}\right) K.
\end{equation}
$G_{sub}$ is the nonrelativistic space group of the
lattice symmetry of the
magnetic sublattices containing elements $[\alpha_R|\tau_\alpha+{\bf R}_n]$.
These operations transform atomic sublattices without
acting on magnetic moments.

In the QS case the symmetry group has the form
\begin{equation}
G_{QS}={\bf C}_L\times G_{tr}^{rel}\times G_{sub}^L
\end{equation}
where $G_{tr}^{rel}=(E,\Theta g_{tr})$ and $G_{sub}^L$
is the subgroup of $G_{sub}$ that leaves invariant the
relativistic terms collinear to {\bf L}.

In the relativistic case the symmetry group has the form
\begin{equation}
G_{rel}=G_{tr}^{rel}\times \mathcal{G}_{sub}^L
\end{equation}
where $\mathcal{G}_{sub}^L$ is the symmetry group of the
magnetic sublattice with operations transforming identically
spin and space subspaces, $\alpha_S$=$\alpha_R$.

The groups $G_{NR}$, $G_{QS}$, and $G_{rel}$ are connected by the
subgroup relation $G_{NR}$$>$$G_{QS}$$>$$G_{rel}$.

Direct consequence of the symmetry with respect to the
subgroup of pure spin rotations about the {\bf L} axis is that all electron states have the spin projection on the
L axis as a good quantum number and that the states with opposite spin projections belong to
different irreducible representations (irreps) of this subgroup \cite{Sandratskii1979}. The latter property guarantees that the states with opposite
spin projections do not hybridize and corresponding bands can intersect with each other.

In the QS case the pure spin rotations about the {\bf L} axis are still the symmetry operations
of the Kohm-Sham Hamiltonian. Therefore all electron states have the spin projection on the
{\bf L} axis as a good quantum number. However, the local atomic operator $l_z$ entering the equation reduces
the number of symmetry operations to those transforming oWe performed systematic calculations for $U$=0.2~Ry (2.72~eV) and some calculations for $U$=0.3~Ry (4.08~eV).
We show the numericperators $l_z$ into themselves.
The complex form
of the SOC operator, in contrast to the real form of the nonrelativistic Hamiltonian, has important
consequences for the formation of atomic orbital moments.

In the relativistic case, separate transformation of the spin and orbital variables becomes
impossible and the SSG reduces itself to a conventional space group of operations $\{\alpha|\alpha|\tau_\alpha+{\bf R}_n\}$. Only
the operations transforming atomic moments into themselves enter the symmetry group of the Kohn-Sham
equation. The pure spin rotations about the Néel vector {\bf L} are not the symmetry operations of the
relativistic problem.

\section{Symmetry-based prediction of magnetic structure instability in ab initio calculations}
\label{App_sec_symWFM}
\renewcommand{\thetable}{C\arabic{table}}
\setcounter{table}{0}

    The method of the prediction is based on the combination of two statements \cite{Sandratskii1996,Sandratskii2001}.
    First: if $g$ is a symmetry
    operation of the initial K-S Hamiltonian this symmetry is preserved during iterations.
    The basis of this statement
    is the fact that the symmetry of the Hamiltonian is reflected in the symmetry properties of the calculated
    electron eigenstates. Since these eigenstates are employed in the modification of the Hamiltonian for the
    next iteration, the modified Hamiltonian is invariant with respect to all
    symmetry operations of the initial Hamiltonian.
    One of the consequences of this statement is that any collinear magnetic structure remains stable in the
    nonrelativistic calculation.

    The property of the symmetry conservation in the DFT calculations is referred to as symmetry constraint \cite{Sandratskii2001}.
    The principle of the symmetry constraint is valid for both nonrelativistic and relativistic calculations.
    However, the symmetry groups and, therefore, the consequences for the same magnetic configuration
    are different.

    The second statement can be seen as an adaptation to the DFT calculations
    of the Dzyaloshinskii's observation that the canting of the atomic moments leading
    to the WFM does not reduce the symmetry of the underlying collinear
    antiferromagnetic state \cite{Dzyaloshinskii1958}.
    This statement says that if the magnetic configuration can be continuously changed without
    destroying any symmetry operation of the initial magnetic configuration, the change {\it must} take place.
    The character of the changes is dictated by the symmetry operations
    of the initial configuration
    that must remain intact according to the statement of symmetry constraint.
    The necessity of the canting is justified as follows.
    Assume that the magnetic configurations we consider are characterized by a continuous parameter $\theta$.
    None of the $\theta$ values is distinguished
    by an additional symmetry and the state with the
    minimal energy cannot be predicted on the symmetry basis. Only direct calculation can give the
    value of $\theta_{min}$. Starting calculation we select an initial value $\theta_0$ of parameter $\theta$,
    often $\theta_0=0$.
    The probability that we accidentally select an unknown $\theta_{min}$ from the continuum of equivalent
    $\theta$ values is zero.
The canting towards the configuration corresponding to the energy minimum must take place.
    In the numerical DFT calculations, the canting is the consequence of nonzero off-diagonal elements
    of the atomic spin-density matrices given by Eq.~(\ref{eq_density_matrix}).
    These elements can be zero for a selected initial magnetic configuration only if there are symmetry
    operations responsible for the zero value. In the WFM case such operations are absent and canting
    takes place.

    In the absence of SOC, both conventional collinear AFM and altermagnets remain stable in the DFT
    calculations for any {\bf L} because
    of the symmetry operations that become lost after any deviation of the
    moments from the initial directions. These are the operations of pure spin rotation about the {\bf L} axis.

    The SOC decreases the symmetry of the problem. The question
    to address is whether the operations protecting the collinearity of the moments of
    the two sublattices remain intact in the relativistic case.
    In AMs the answer depends on the
    direction of the {\bf L} vector with respect to the crystal lattice.

\begin{table}
\caption{Symmetry operations of an antiferromagnet with NiAs crystal lattice
and Néel vector collinear to the $x$ axis. Relativistic consideration.
First column: number of the operation. Second column: symbol of the operation.
$E$ is the unity operation; $C_{2\alpha}$ are rotations
by angle 180$^{\circ}$ about
$\alpha$ axis, $\alpha=x,y,z$;  $\sigma_\beta$ is the reflection in the $\beta$ plane,
$\beta=x,y,z$ correspond to the $x=0$, $y=0$, $z=0$ planes, $\Theta$ is the operation of time reversal.
Third column: if non-zero, gives the vector of nonprimitive translation $\mathbf{\bm{\tau}}=(0,0,0.5)$.
Forth column: the result of the action of the point operation on axial vector $( m_x, m_y, m_z)$.
Fifth column: the atoms into which the atoms Ni$_1$, Ni$_2$, As$_1$, As$_2$ are transformed by the
operation.
}
\begin{ruledtabular}
\begin{tabular}{lcccc}
No.&OP         &  $\mathbf{\bm{\tau}}$  & $\mathbf{m}$ & at. permutation\\
\hline\\
1  &$E$           & $0$ &      $( m_x, m_y, m_z)$&         Ni$_1$Ni$_2$$\:$ As$_1$As$_2$\\
2  &$\Theta$$C_{2x}$      & $\bm{\tau}$ &   $(m_x,-m_y, -m_z)$&    Ni$_2$Ni$_1$$\:$ As$_1$As$_2$\\
3  &$\Theta$$C_{2y}$      & $0$ &      $(-m_x,m_y, -m_z)$&         Ni$_1$Ni$_2$$\:$ As$_2$As$_1$\\
4  &$C_{2z}$      & $\bm{\tau}$ &      $(-m_x,-m_y, m_z)$& Ni$_2$Ni$_1$$\:$ As$_2$As$_1$ \\
5  &$I$           & $0$ &      $( m_x, m_y, m_z)$&         Ni$_1$Ni$_2$$\:$ As$_2$As$_1$\\
6  &$\Theta$$\sigma_x$   & $\bm{\tau}$ &    $( m_x, -m_y, -m_z)$&  Ni$_2$Ni$_1$$\:$ As$_2$As$_1$\\
7  &$\Theta$$\sigma_y$   & $0$ &    $( -m_x,m_y, -m_z)$&           Ni$_1$Ni$_2$$\:$ As$_1$As$_2$\\
8  &$\sigma_z$    & $\bm{\tau}$ &      $(-m_x,-m_y, m_z)$& Ni$_2$Ni$_1$$\:$ As$_1$As$_2$\\
\end{tabular}
\end{ruledtabular}
\label{tab_sym}
\end{table}
Let us consider the relativistic case of ${\bf L}||x$. The eight point symmetry transformations
are collected in Table~\ref{tab_sym}.
For Mn and Te atomic moments of MnTe we get the following constraints:\\
$\begin{array}{lll}
m_x({\rm Mn}_1)=-m_x({\rm Mn}_2)  &\:\:\:&  m_x({\rm Te}_1)=m_x({\rm Te}_2)=0 \\
m_y({\rm Mn}_1)=m_y({\rm Mn}_2)=0 &\:&  m_y({\rm Te}_1)=m_y({\rm Te}_2)=0 \\
m_z({\rm Mn}_1)=m_z({\rm Mn}_2)   &\:&  m_z({\rm Te}_1)=m_z({\rm Te}_2).\\
\end{array}$
We see that the zero values of the $z$-componets of the Mn and Te atomic moments are not
among the constraints imposed by the symmetry operations.
Accordingly, the appearance of the $z$ components
of the Mn and Te moments will not destroy any of the symmetry operations of the initial magnetic
configuration. Indeed, the iterations immediately result in the canting of the Mn moments towards the
$z$ axis and the appearance of the Te moments collinear to the $z$ axis. In the relativistic calculation,
the symmetry conditions on the atomic spin and orbital moments are identical. Therefore, the calculations
give canted Mn orbital moments and orbital moments of Te atoms collinear to the hexagonal $c$ axis.

In the case of ${\bf L}||y$ there are the same 8 unitary symmetry operations but, in contrast to ${\bf L}||x$ case,
all operations are without time reversal.
We get the following constraints\\
$\begin{array}{lll}
m_x({\rm Mn}_1)=m_x({\rm Mn}_2)=0  &\:\:\:&  m_x({\rm Te}_1)=m_x({\rm Te}_2)=0 \\
m_y({\rm Mn}_1)=-m_y({\rm Mn}_2) &\:&  m_y({\rm Te}_1)=m_y({\rm Te}_2)=0 \\
m_z({\rm Mn}_1)=m_z({\rm Mn}_2)=0   &\:&  m_z({\rm Te}_1)=m_z({\rm Te}_2=0)\\
\end{array}$
that forbid the deviation of the Mn moments from the $y$ axis and nonzero values of the Te moments.

For completeness, we remark that if the initial direction of the Néel vector lie
in the $xy$ plane, intermediate between $x$ and $y$ axes, the symmetry predicts the necessity of the rotation of the
L vector in the $xy$ plane and, simultaneously, the canting of the Mn moments towards the $z$ axis.
On the Te atoms appear magnetic moments collinear to the $z$ axis.
If the initial direction of {\bf L} is
a general direction in space, all symmetry constraints are lost. In this case the pairs of the
Mn and Te atoms become inequivalent and iterations lead to different values and directions of the atomic moments
in these pairs.
Such magnetic configurations
can be characterized as the combination of weak ferromagnetic and weak ferrimagnetic \cite{Carmine2025}
distortions of the initial collinear configuration. Since there are no constraints, no symmetry-based prediction
can be made about the properties of the magnetic configuration obtained as the result of the convergence
of the iteration process.

The analysis of the quasisymmetry calculation gives a remarkable result. Here the relativistic terms
has the form $\sim$$\sigma_L\hat{l}_L$ where we took into account that local atomic $z$ axes are collinear
to the Néel vector {\bf L}. This operator commutes with spin rotations about the {\bf L} axis. Therefore the
spin projection of the electron states on the {\bf L} axis remains a good quantum number and the directions
of the atomic spin moments remain collinear to the {\bf L} axis during iterations. However, the presence of
operator $\hat{l}_L$ leads to the formation of the orbital atomic moments whose symmetry properties are identical
to the symmetry properties of the orbital moments in the relativistic calculation with full account for
the SOC.

\section{Magnetic moments as functions of the SOC strength}
\subsection{Relativistic calculations}
\label{App_sec_moments_of_gamma}

\begin{figure}[h]
\includegraphics*[width=8cm]{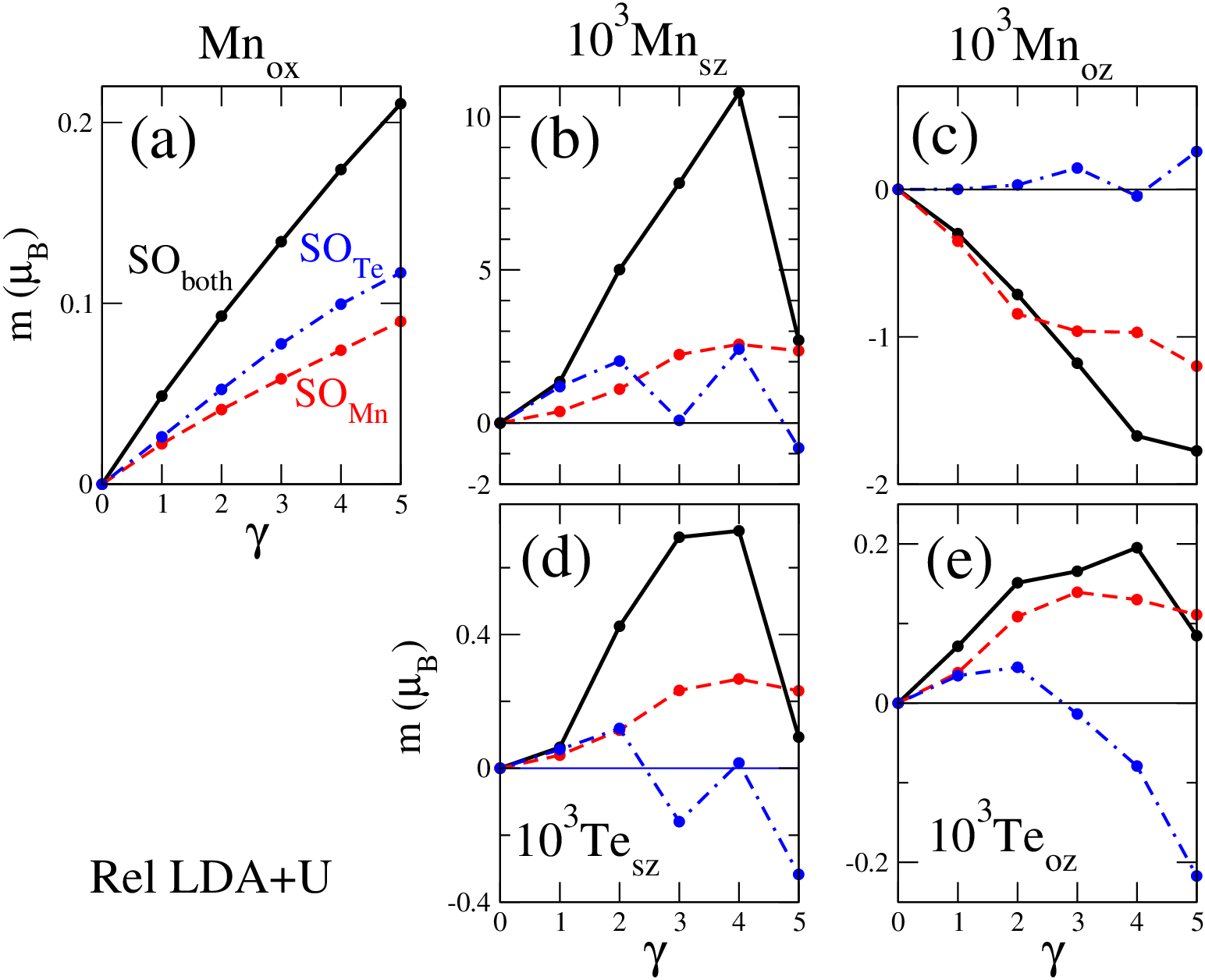}
\caption{The same as in Fig.~\ref{fig_mom_of_SOC_MnTe_LDA} but calculated with LDA+$U$ method, $U=$0.2~Ry.}
\label{fig_mom_of_SOC_MnTe_LDA+U}
\end{figure}

\begin{figure}[h]
\includegraphics*[width=8cm]{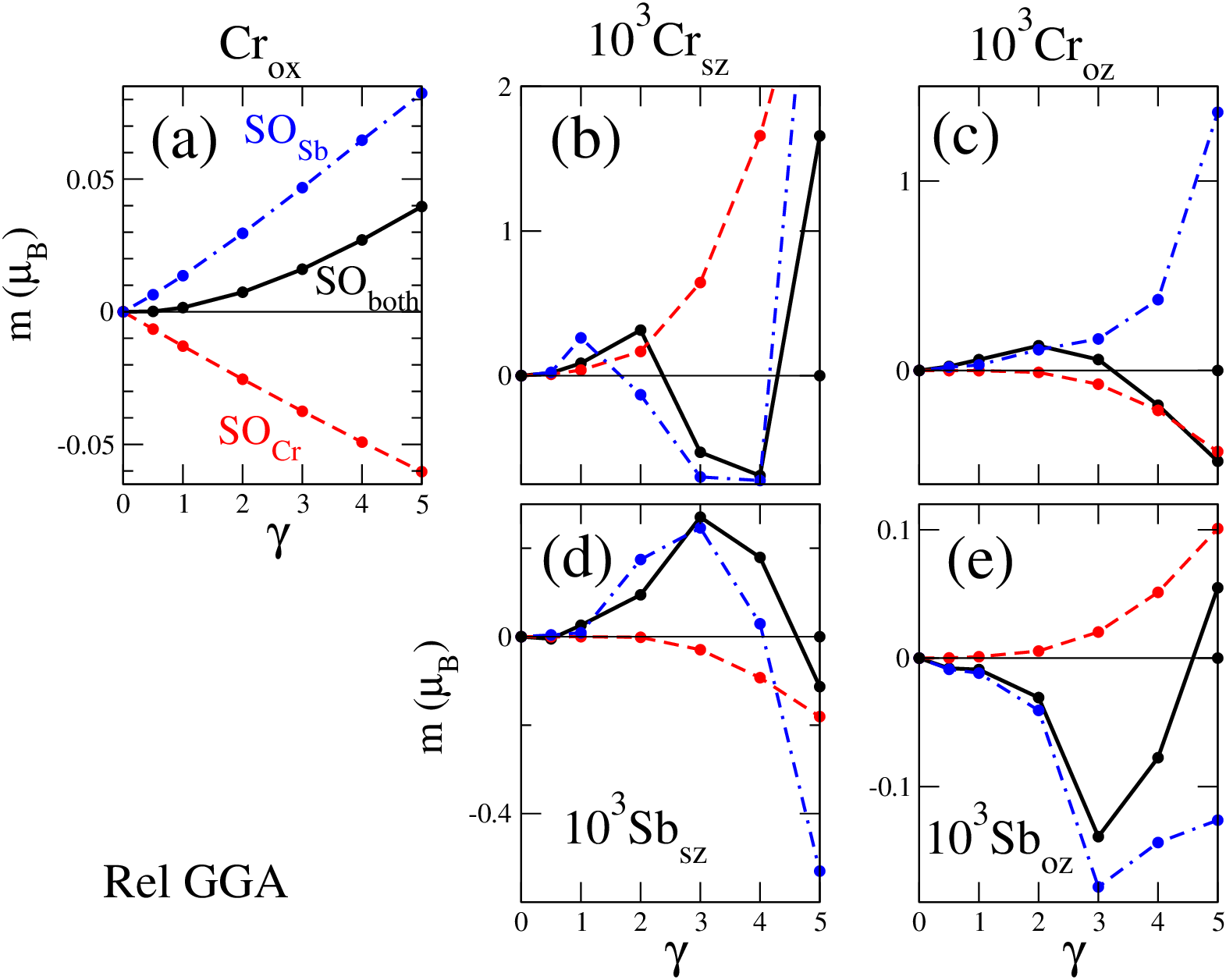}
\caption{The same as in Fig.~\ref{fig_mom_of_SOC_MnTe_LDA} but calculated with GGA potential for CrSb.}
\label{fig_mom_of_SOC_CrSb_GGA}
\end{figure}

The {\bf L} component of the orbital Cr moment induced by the Cr-SOC in CrSb is negative [Fig.~\ref{fig_mom_of_SOC_CrSb_GGA}(a)],
in contrast to positive
orbital Mn moment induced by Mn-SOC in MnTe [Figs.~\ref{fig_mom_of_SOC_MnTe_LDA}~and~\ref{fig_mom_of_SOC_MnTe_LDA+U}].
The orbital moments induced by the Lg-SOC are positive
for both materials. As a consequence, in CrSb the Cr-SOC and Sb-SOC contributions strongly
compensate each other when both SOCs are switched on while in MnTe the total-SOC value is
larger than both partial-SOC values.

\subsection{Quasisymmetry calculation}
\label{sec_QS_moment}

\begin{figure}[h]
\includegraphics*[width=8cm]{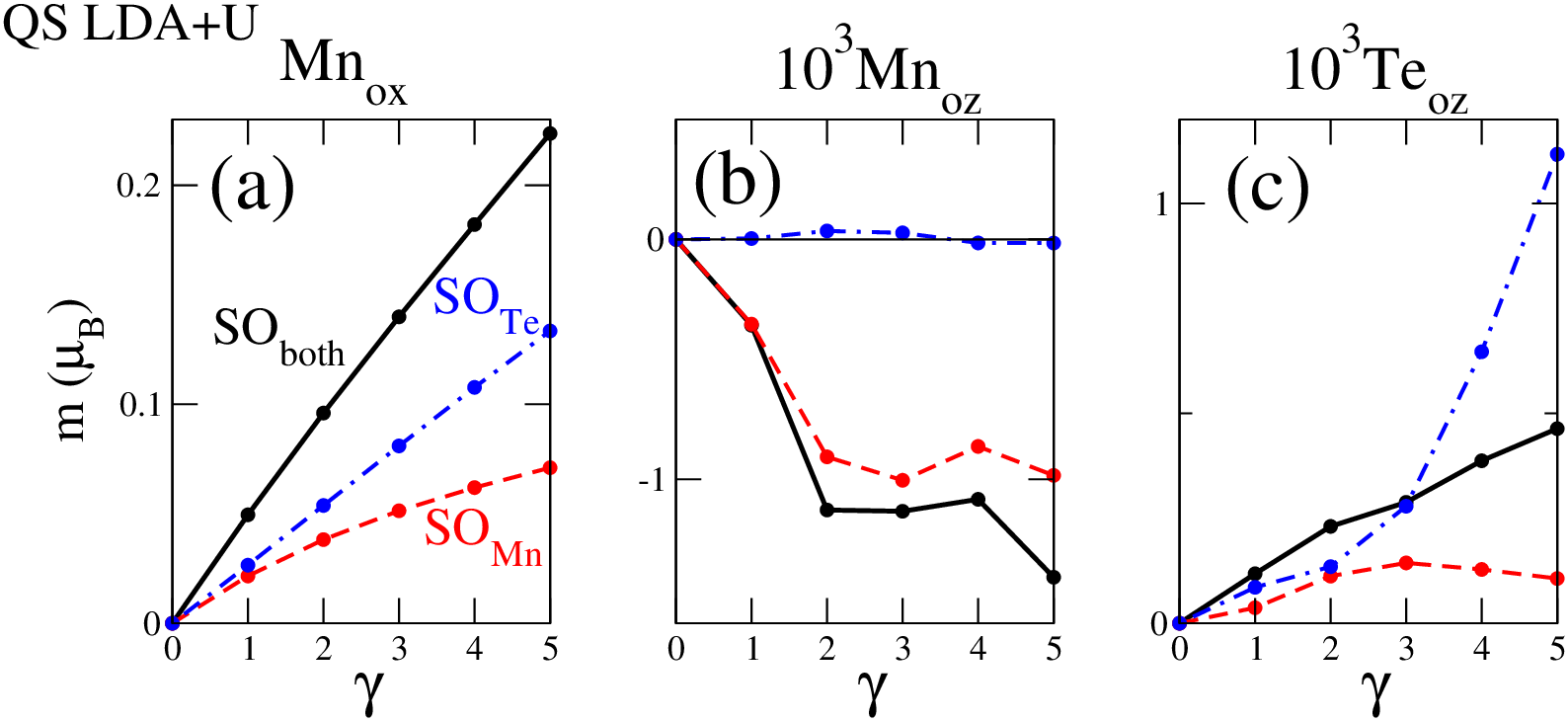}
\caption{Induced moments as functions of the
SOC strength. LDA+$U$ quasisymmetry calculation for MnTe.
(a) The component of the Mn orbital moment collinear to the Néel vector.
(b) and (c) The components of the Mn and Te orbital moments
collinear to the $z$ axis and contributing
to WFM. These components are multiplied by $10^3$. Colors of the curves are the same as
in Fig.~\ref{fig_mom_of_SOC_MnTe_LDA}.}
\label{fig_mom_of_SOC_MnTe_LDA+U_QS}
\end{figure}

\begin{figure}[h]
\includegraphics*[width=8cm]{ 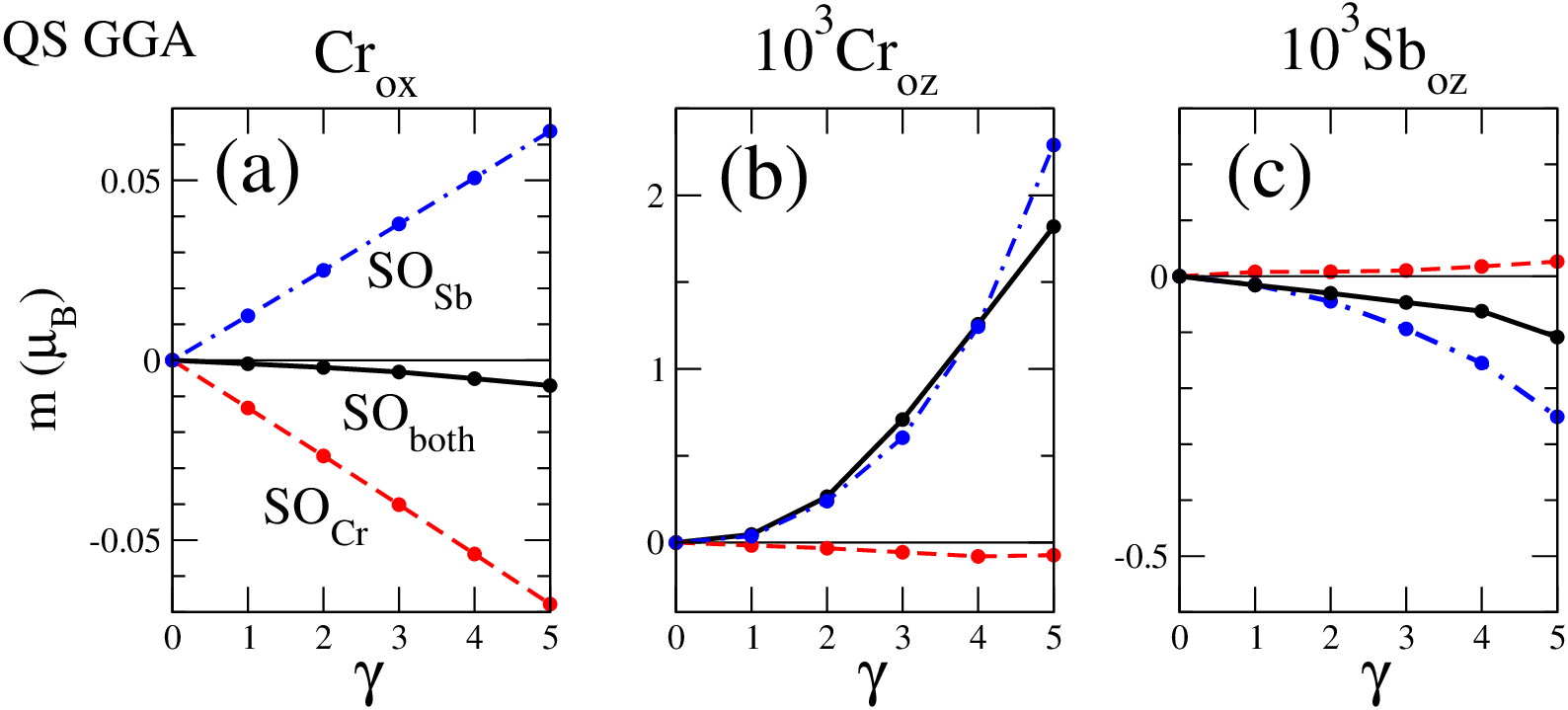}
\caption{Induced moments as functions of the SOC strength. GGA quasisymmetry calculation for CrSb.
(a) The component of the Cr orbital moment collinear to the Néel vector.
(b) and (c) The components of the Cr and Sb orbital moments
collinear to the $z$ axis and contributing
to WFM. These components are multiplied by $10^3$. Colors of the curves are the same as in
in Fig.~\ref{fig_mom_of_SOC_MnTe_LDA}}
\label{fig_mom_of_SOC_CrSb_GGA_QS}
\end{figure}

The quasisymmetry calculation for MnTe [Fig.~\ref{fig_mom_of_SOC_MnTe_LDA+U_QS}(a)] shows that the SOC component
collinear to {\bf L} is to a
large extent responsible for the {\bf L} component of the Mn orbital moment [Fig.~\ref{fig_mom_of_SOC_MnTe_LDA+U}(a)].
In the case of CrSb, the situation is more peculiar. Here the $\gamma$ dependencies of the {\bf L} component of the
Cr moment obtained with switching off the SOC on one of the atoms are rather similar in the quasisymemtry and relativistic
calculations [Figs.~\ref{fig_mom_of_SOC_CrSb_GGA}(a)~and~\ref{fig_mom_of_SOC_CrSb_GGA_QS}(a)]. However, because of the
strong compensation of partial contributions, the dependencies obtained in the calculations with account for
the SOC contributions of both atomic types are quantitatively noticeably different.


\section{The LDA+U calculations of band structure and MSES. }
\label{App_sec_LDA+U}

\begin{figure}[h]
\includegraphics*[width=8cm]{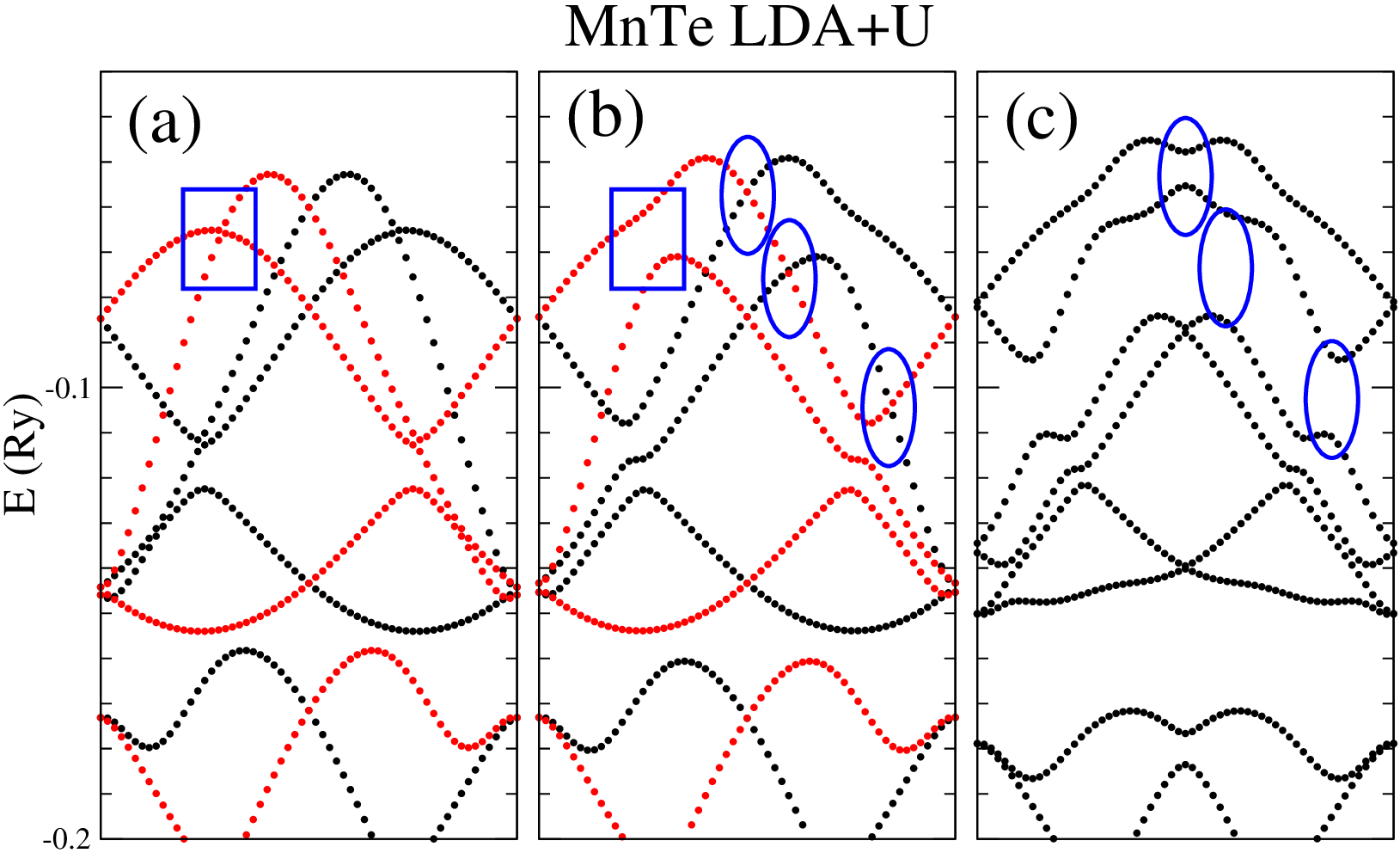}
\caption{The same as in Fig.~\ref{fig_Bands_LDA} but for LDA+$U$ calculation.}
\label{fig_Bands_LDA+U}
\end{figure}

\begin{figure}[h]
\includegraphics*[width=8cm]{ 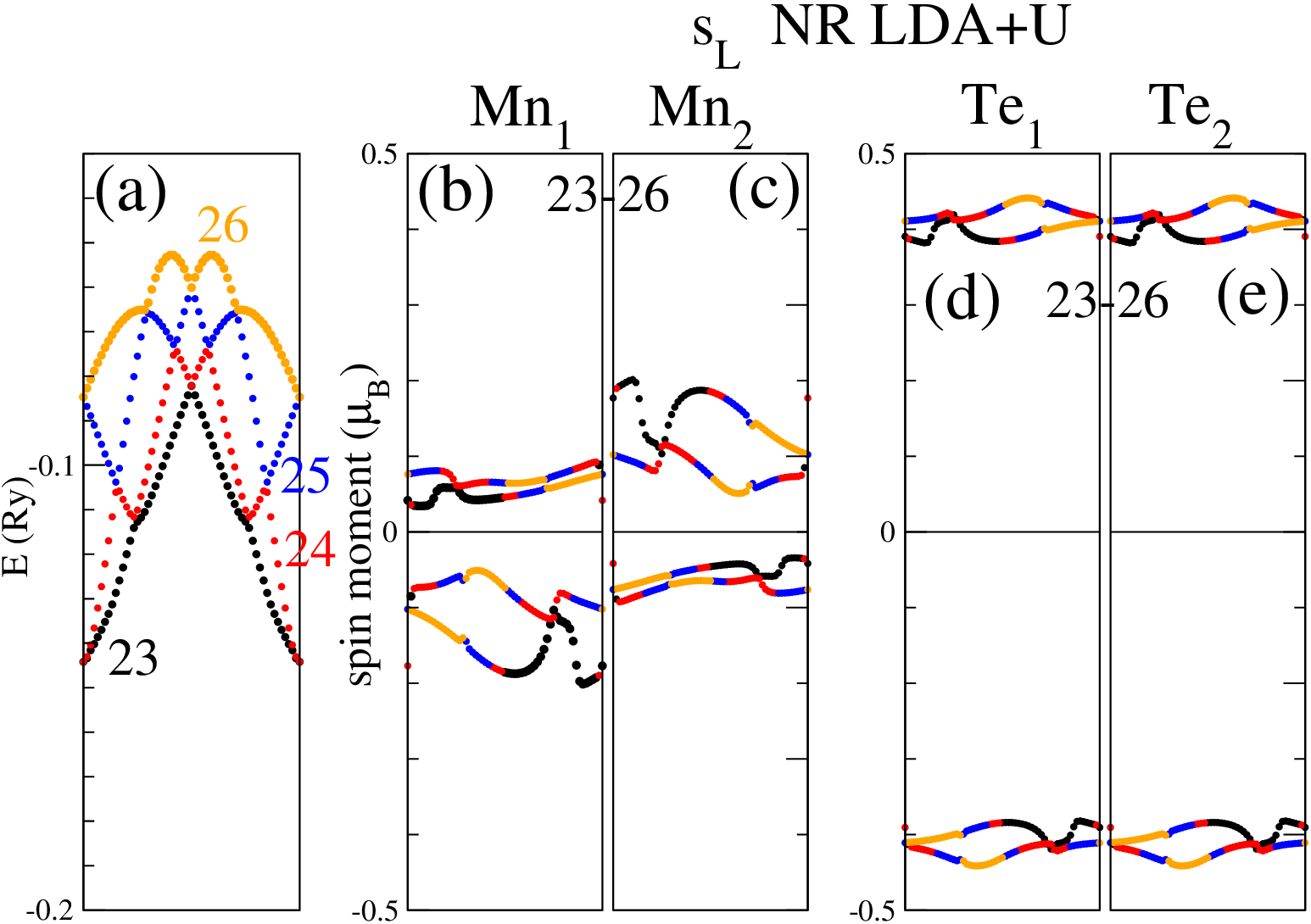}
\caption{The same as in Fig.~\ref{fig_MSES_SO0_LDA} but calculated with LDA+$U$.}
\label{fig_MSES_SO0_LDA+U}
\end{figure}
\begin{figure}[h]
\includegraphics*[width=8cm]{ 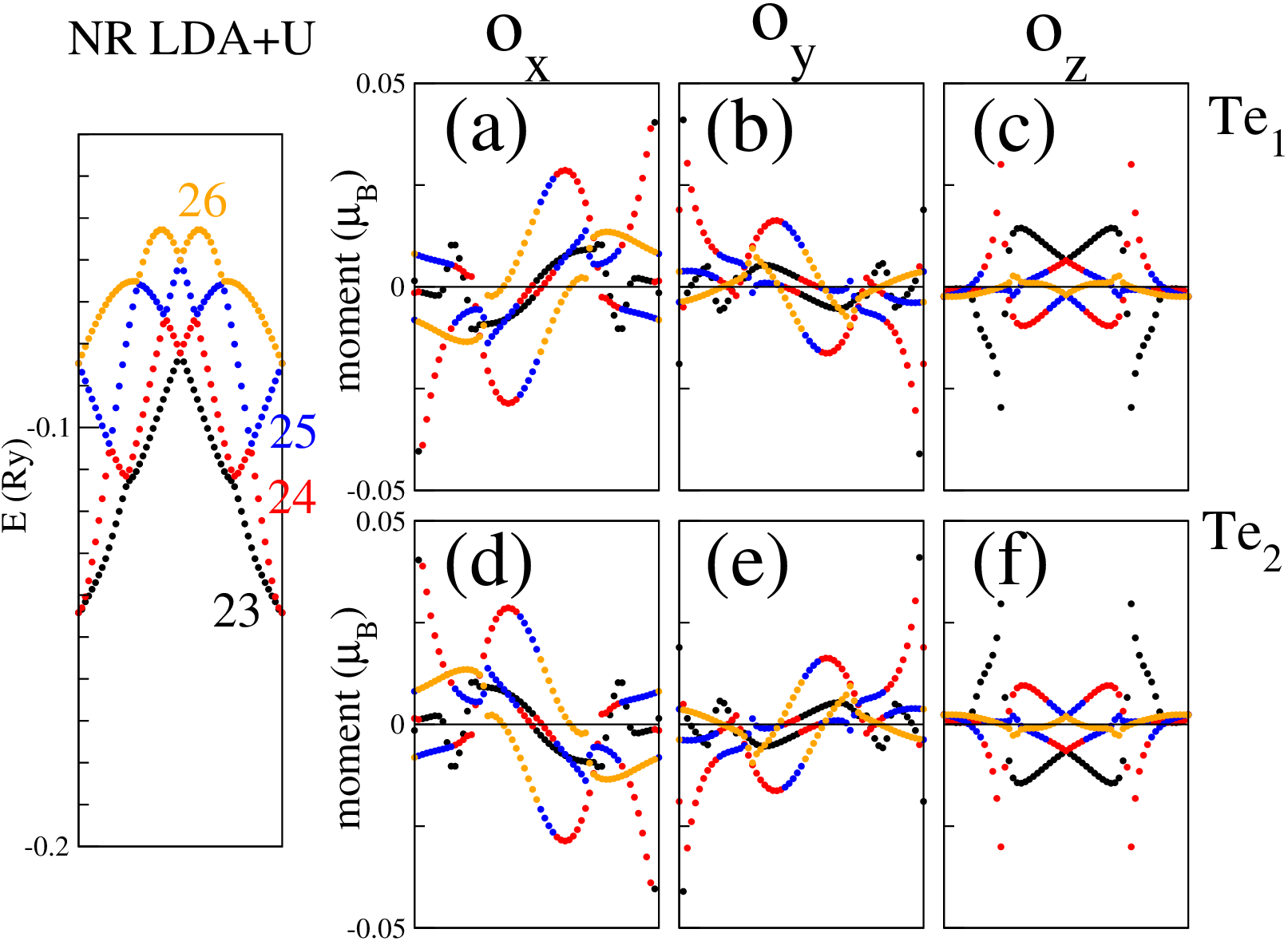}
\caption{Nonrelativistic orbital-MSES of MnTe calculated with LDA+$U$.
Shown are the $x$, $y$, and $z$ projections of the orbital moments of the Te$_1$
[panels (a)-(c)] and Te$_2$ [panels (d)-(f)] atoms.}
\label{fig_ORBIT-MSES_SO0_LDA+U}
\end{figure}
\begin{figure}[h]
\includegraphics*[width=8cm]{ 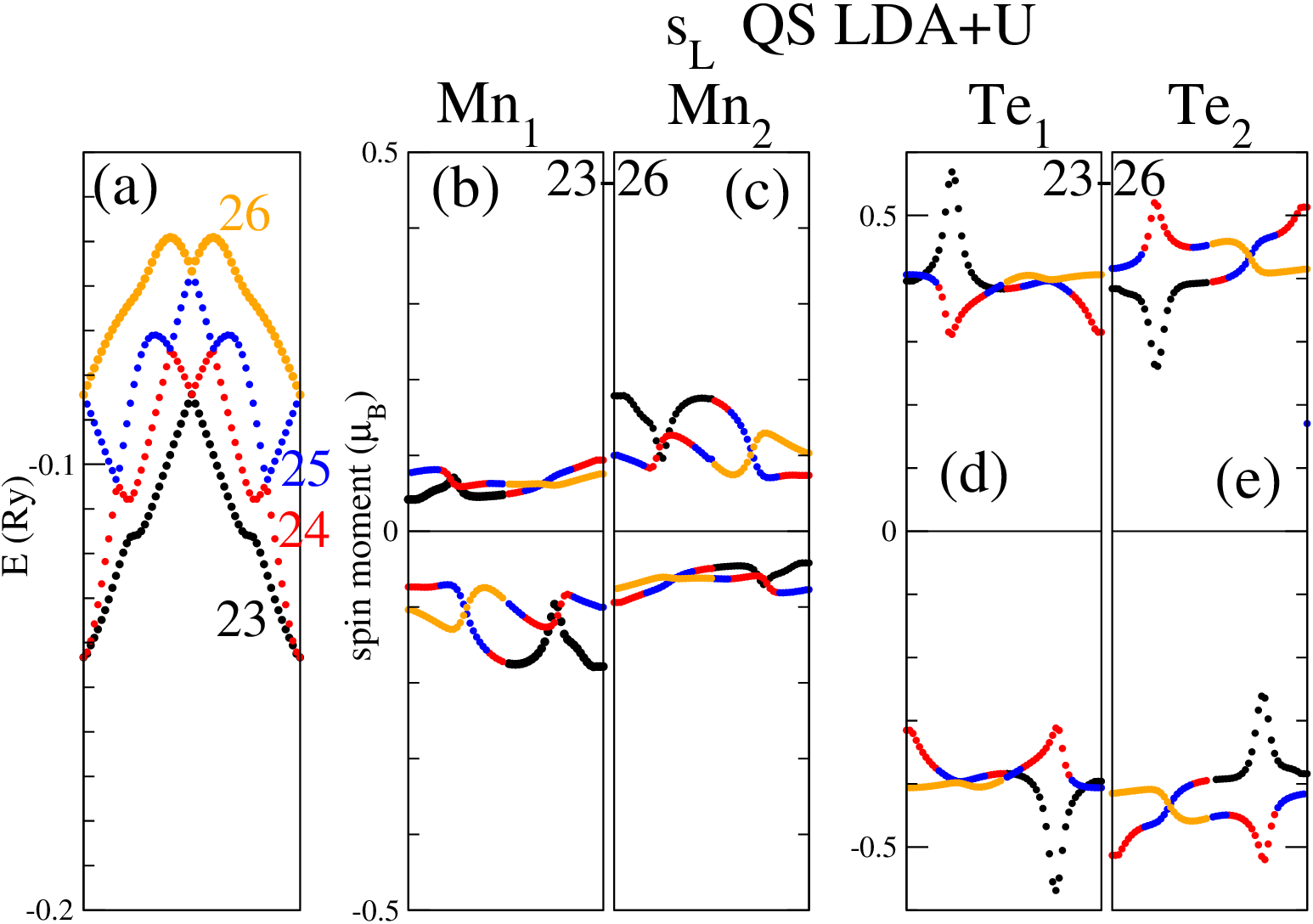}
\caption{The same as in Fig.~\ref{fig_MSES_SO0_LDA} but calculated with quasisymmetry LDA+$U$.}
\label{fig_spin-MSES_SO1X_LDA+U}
\end{figure}
\begin{figure}[t]
\includegraphics*[width=8cm]{ 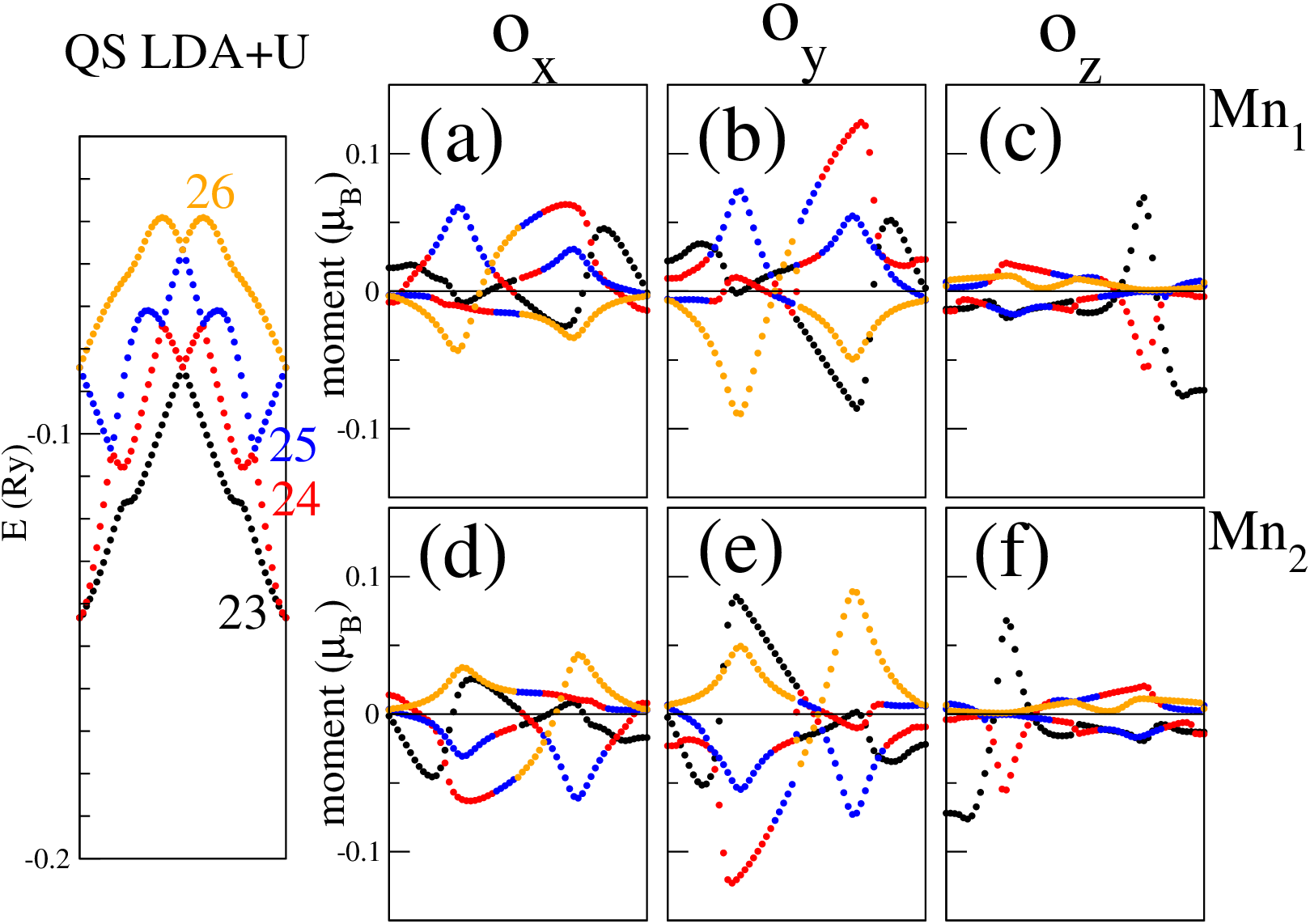}
\caption{Orbital-MSES of MnTe calculated with quasisymmetry LDA+$U$.
Shown are the $x$, $y$, and $z$ projections of the orbital moments of the Mn$_1$ [panels (a)-(c)] and Mn$_2$ [panels (d)-(f)] atoms.}
\label{fig_orbital-Mn-MSES_SO1X_LDA+U}
\end{figure}
\begin{figure}[h]
\includegraphics*[width=8cm]{ 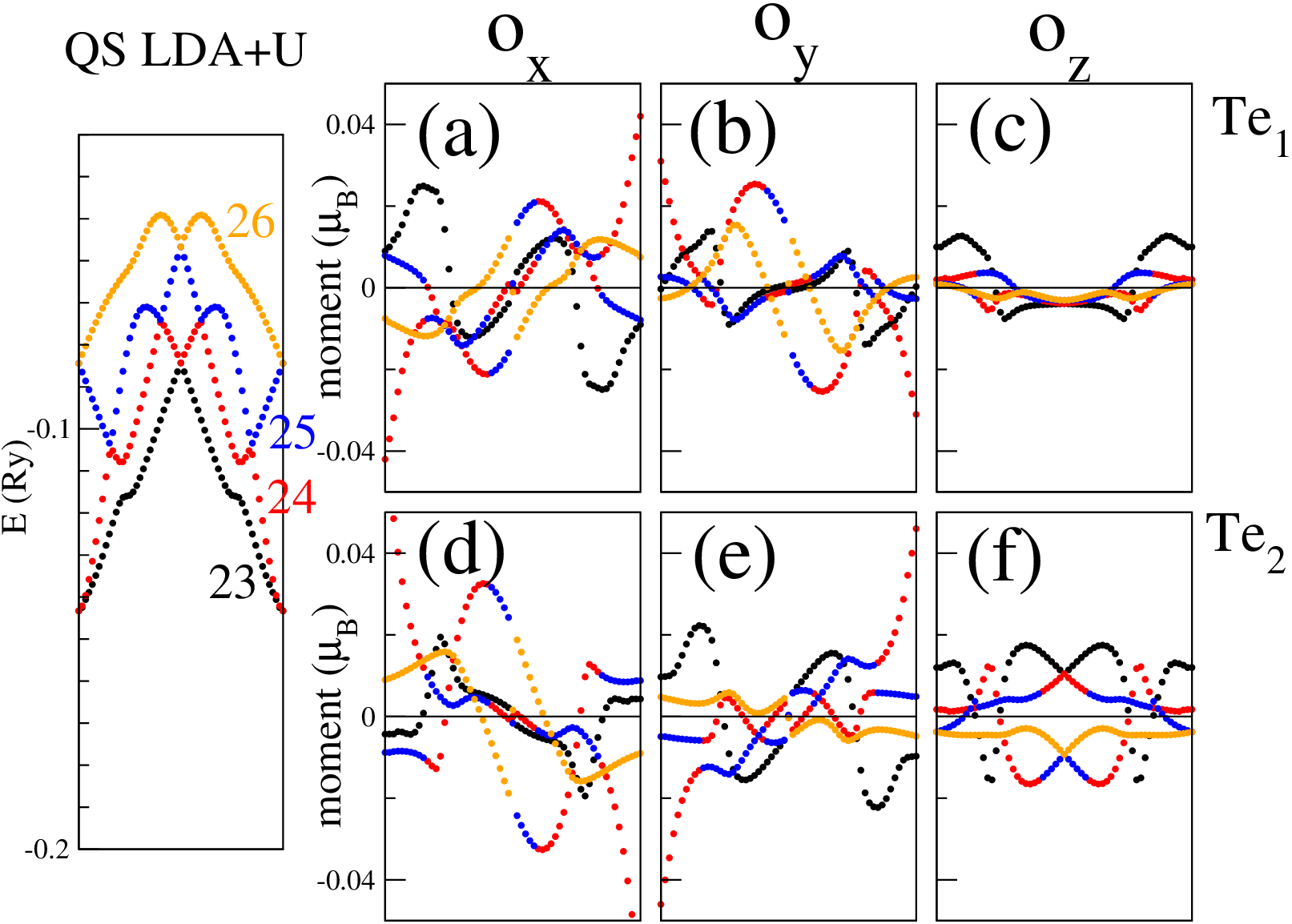}
\caption{The same as in Fig.~\ref{fig_orbital-Mn-MSES_SO1X_LDA+U} but for the Te$_1$ and Te$_2$ atoms.}
\label{fig_orbital-Te-MSES_SO1X_LDA+U}
\end{figure}
\begin{figure}[h]
\includegraphics*[width=8cm]{ 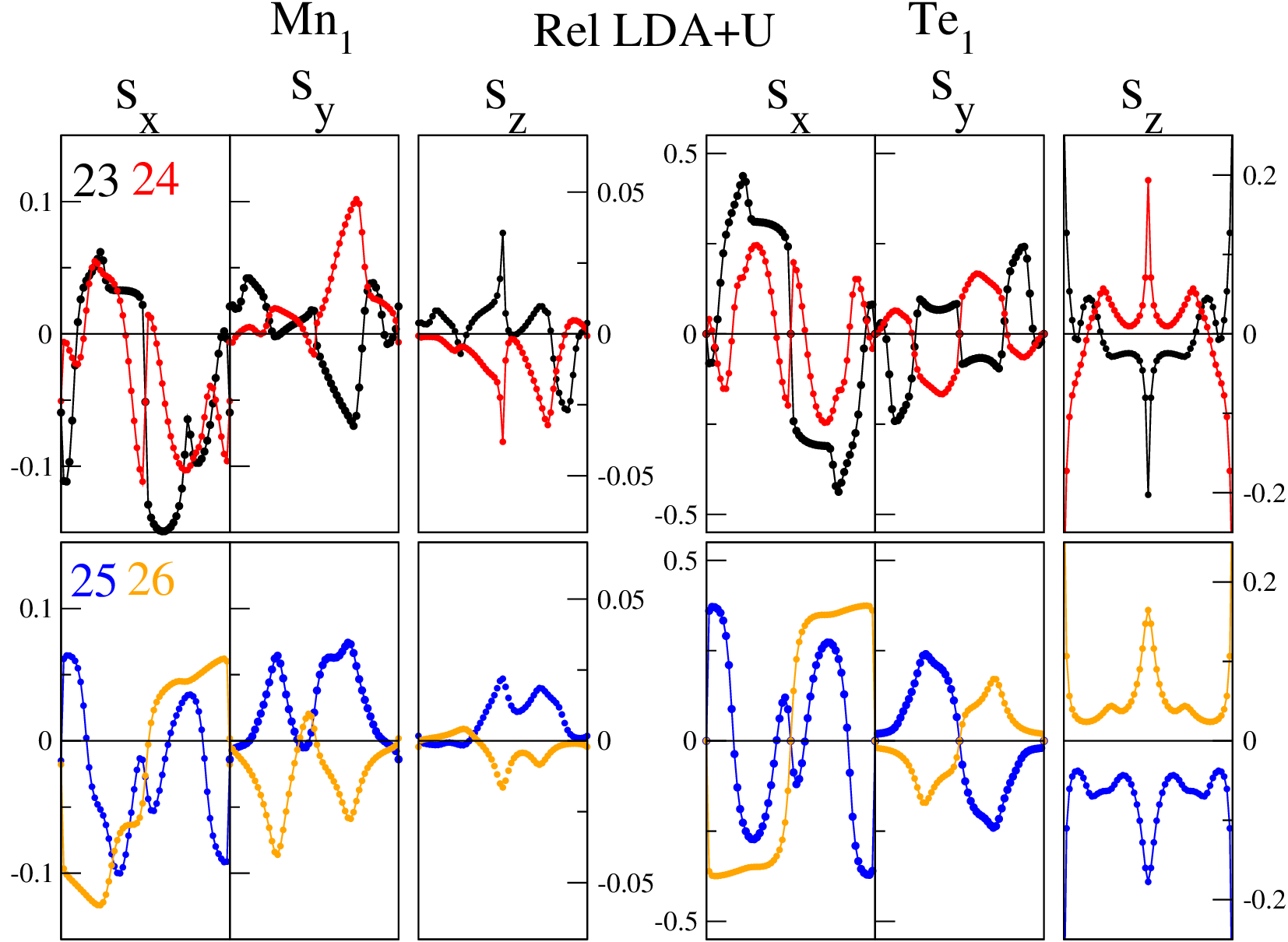}
\caption{The same as in Fig.~\ref{fig_spin-MSES_SO1_LDA} but calculated with LDA+$U$.}
\label{fig_spin-MSES_SO1_LDA+U}
\end{figure}
\begin{figure}[h]
\includegraphics*[width=8cm]{ 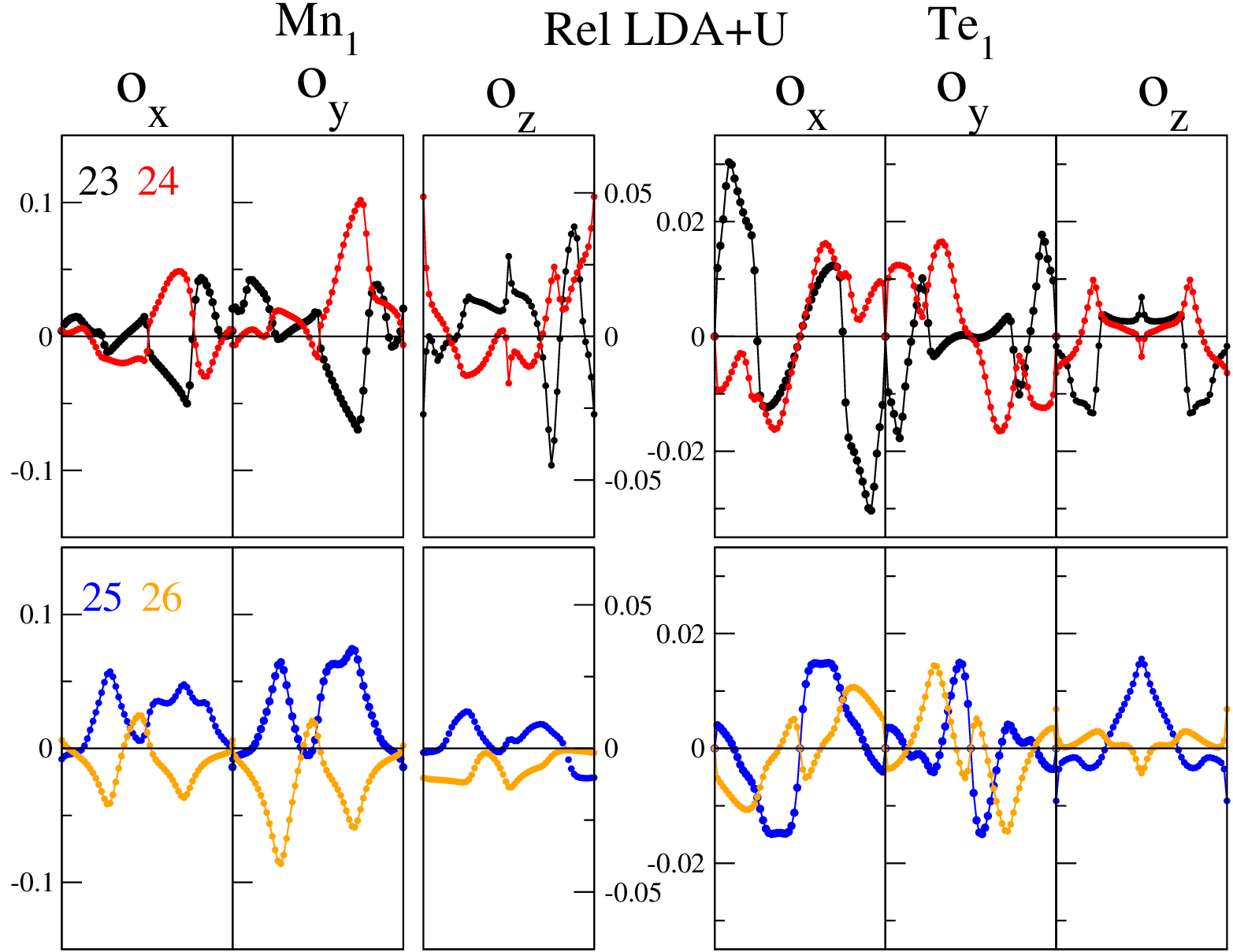}
\caption{The same as in Fig.~\ref{fig_spin-MSES_SO1_LDA} but showing orbital-MSES
instead of spin-MSES and calculated with LDA+$U$ instead of LDA. }
\label{fig_orbital-MSES_SO1_LDA+U}
\end{figure}

\end{document}